\documentclass[prx,twocolumn,aps]{revtex4-2}
\usepackage{amsmath,bm,epsfig,color}

\usepackage{xcolor,hyperref}
\hypersetup{
   colorlinks,
   linkcolor={blue!50!black},%{red!80!black},
   citecolor={blue!50!black},
   urlcolor={blue!80!black}
}
\usepackage[utf8]{inputenc}
\usepackage{graphicx}

\renewcommand{\=}{\!=\!}
\usepackage[]{inputenc}
\graphicspath{{figs/}}

\newcommand{\Vstar}{v_*}

\begin{document}

\title{A finite geometry, inertia assisted coarsening-to-complexity transition\\ in homogeneous frictional systems}
\author{Thibault Roch$^{1}$}
\author{Efim A.~Brener$^{2,3}$}
\author{Jean-Fran\c{c}ois Molinari$^{1}$}
\thanks{jean-francois.molinari@epfl.ch}
\author{Eran Bouchbinder$^{4}$}
\thanks{eran.bouchbinder@weizmann.ac.il}
\affiliation{$^{1}$Civil Engineering Institute, Materials Science and Engineering Institute, Ecole Polytechnique F\'ed\'erale de Lausanne, Station 18, CH-1015 Lausanne, Switzerland\\
$^{2}$Peter Gr\"unberg Institut, Forschungszentrum J\"ulich, D-52425 J\"ulich, Germany\\
$^{3}$Institute for Energy and Climate Research, Forschungszentrum J\"ulich, D-52425 J\"ulich, Germany\\
$^{4}$Chemical and Biological Physics Department, Weizmann Institute of Science, Rehovot 7610001, Israel}
%\date{\today}

\begin{abstract}
The emergence of statistical complexity in frictional systems (where nonlinearity and dissipation are confined to an interface), manifested in broad distributions of various observables, is not yet understood. We study this problem in velocity-driven, homogeneous (no quenched disorder) unstable frictional systems of height $H$. The latter are described at the continuum scale within a realistic rate-and-state friction interfacial constitutive framework, where elasto-frictional instabilities emerge from rate-weakening friction. For large $H$, such frictional systems were recently shown to undergo continuous coarsening until settling into a spatially periodic traveling solution. We show that when the system's height-to-length ratio becomes small --- characteristic of various engineering and geophysical systems ---, coarsening is less effective and the periodic solution is dynamically avoided. Instead, and consistently with previous reports, the system settles into a stochastic, statistically stationary state. The latter features slip bursts, whose slip rate is larger than the driving velocity, which are non-trivially distributed. The slip bursts are classified into two types: predominantly non-propagating, accompanied by small total slip and propagating, accompanied by large total slip. The statistical distributions emerge from dynamically self-generated heterogeneity, where both the non-equilibrium history of the interface and wave reflections from finite boundaries, mediated by material inertia, play central roles. Specifically, the dynamics and statistics of large bursts reveal a timescale $\sim\!H/c_{\rm s}$, where $c_{\rm s}$ is the shear wave-speed. We discuss the robustness of our findings against variations of the frictional parameters, most notably affecting the magnitude of frictional rate-weakening, as well as against different interfacial state evolution laws. Finally, we demonstrate a reverse transition in which statistical complexity disappears in favor of the spatially periodic traveling solution. Overall, our results elucidate how relatively simple physical ingredients can give rise to the emergence of slip complexity.
\end{abstract}

\maketitle

%\subsection*{Introduction}

Complexity in physical systems, i.e., the existence of broad, fat-tailed distributions of various observables, can emerge from quenched disorder~\cite{hansen1991roughness,nattermann1992dynamics,fisher1998collective,sethna2001crackling,zaiser2006scale,stauffer2018introduction}, from sufficiently strong bulk nonlinearity (e.g, as in turbulence~\cite{frisch1995turbulence}) and/or due to the spontaneous emergence of a critical state (e.g, as in self-organized criticality~\cite{bak1988self}). Whether complexity might emerge even in their absence is not yet clarified and more generally, we still lack a complete understanding of the minimal conditions for the emergence of complexity.

Frictional systems, typically composed of two deformable bodies that interact under compressive forces along a contact interface, offer a rich test bed for addressing these questions. This is the case because such systems feature generic physical ingredients that appear relevant, including nonlinearity and dissipation --- mostly confined to the frictional interface, i.e., a low-dimensional object embedded in larger dimensional ones (the surrounding deformable bulk materials) ---, long-range interactions mediated by bulk deformation, material inertia, elasto-frictional instabilities and various forms of disorder.

The complexity-related interest in frictional systems has a long history that is not exclusively (or even mainly) motivated by statistical physics questions, but rather by geophysical observations. It is well established that the statistical properties of earthquakes --- i.e., frictional rupture --- occurring on natural faults reveal complexity~\cite{Carlson1994,Rundle2003,Kawamura2012}. That is, many earthquake-related observables, such as the magnitude of earthquakes~\cite{Gutenberg1944,Utsu1999}, the number of aftershocks as a function of time following a main earthquake~\cite{Omori1894,Utsu1995} and various other slip-related physical quantities~\cite{Kanamori1975,Davidsen2004,Wells1994,Scholz2019}, are known to be power-law distributed, oftentimes with apparently universal exponents.

These observations have driven extensive efforts aiming at revealing the physical origins of complexity. Among the many research avenues taken to shed light on this problem, which we cannot possibly review here (see, for example,~\cite{Carlson1989b,Carlson1989a,Carlson1991,Langer1996,Shaw1994,Nielsen2000,Shaw1997,Myers1996}), we focus here on the one that considers coarse-grained interfacial constitutive relations with a well-defined continuum limit and homogeneous systems~\cite{Rice1993,Ben-zion1993,Ben-zion1995,Rice1996,ben1997dynamic,Ben-zion2001,cattania2019complex}. The latter implies that the deformable bodies that form the frictional interface are homogeneous and isotropic, typically described by linear elastodynamics, that the interface is planar on the continuum scale and that the frictional properties feature no spatial variability or quenched disorder. In the simplest scenario, the two bodies are assumed to be two-dimensional (2D) and identical, each featuring height $H$ and length $L$ (in the sliding direction). Frictional sliding is driven by applying a relative velocity $v_0$ at the outer boundaries.

The interfacial constitutive framework considered is based on realistic, laboratory-derived friction laws that incorporate both the rate dependence of the frictional strength and its dependence on the structural state of the interface~\cite{Dieterich1978,Ruina1983,Rice1983,Dieterich1986,Tullis1986,Kilgore1993,Dieterich1994,Marone1998a,Baumberger1999,Nakatani2001,Rubinstein2004,Baumberger2006,Dieterich2007,Ben-david2010,Reches2010,Nagata2012,Bhattacharya2014,Barsinai2014,Rubino2017}. The latter typically corresponds to the real contact area~\cite{Dieterich1994,Baumberger2006,Bowden2001}, which is a function of the contact age, taken as the frictional state field that encodes memory of the history of sliding. Within this interfacial constitutive framework, once coupled to the elasticity of the surrounding bulks, the system features a linear elasto-frictional instability with a minimal wavelength $L_{\rm c}\!<\!L$ over a range of homogeneous sliding velocities~\cite{Rice1983,Rice1993,Aldam2017}. The latter property implies that homogeneous sliding at a velocity $v_0$ is not possible, if $v_0$ is within the unstable velocities range. Hence, if the system settles into a stationary state, it must be spatiotemporally inhomogeneous (long sliding times are typically attained by employing periodic boundary conditions in the sliding direction).

The question then boils down to whether the inhomogeneous state reveals nontrivial aspects of complexity in various physical observables. For the latter to occur, in the absence of quenched disorder and/or geometric irregularities, self-generated heterogeneity in the interfacial stress and state fields should spontaneously emerge, and feature strong enough fluctuations that can arrest propagating frictional rupture over a broad range of scales.

Models of crustal faults, employing the above-described continuum rate-and-state constitutive framework and typically formulated in 3D, indicated that significant complexity does not generically emerges in the limit $H/L\!\gg\!1$~\cite{Rice1993,Ben-zion1993,Ben-zion1995,Rice1996,ben1997dynamic,Ben-zion2001}. That is, while some aspects of complexity do emerge for large $H/L$ ratios, their existence might depend on the interfacial state evolution law and/or on the interfacial constitutive parameters~\cite{Rice1996}. These results suggest that in the large $H/L$ limit, the growing stress concentration near the edges of frictional rupture is so large that the self-generated heterogeneity is not generically strong enough to arrest rupture. In such cases, the long-time behavior of the system is characterized by quasi-periodic slip bursts that correspond to rupture propagating a distance comparable to $L$~\cite{Rice1996}.

The latter has been recently demonstrated in the $H/L\!\to\!\infty$ limit for strictly 2D frictional systems that feature no spatial variation in the frictional properties~\cite{Roch2022}. In particular, it has been shown that such systems feature complex dynamics over finite times, but in fact undergo continuous coarsening dynamics that are terminated on the scale of the system length $L$. That is, in the long-time limit, the system settles into a deterministic, regular steady state in which a single self-healing slip pulse steadily propagates through the periodic boundary conditions~\cite{Roch2022}. We will use in this work the results of~\cite{Roch2022} as a reference case for the absence of generic complexity in the large $H/L$ limit, being advantageous for our purposes compared to the crustal fault models mentioned above that feature some depth dependence of some properties (e.g., frictional or in the boundary conditions) and their numerical analysis usually involves various approximations (e.g., an ad hoc 2D approximation to the 3D problem or the quasi-dynamic approximation to elastodynamics in 3D)~\cite{Rice1996,cattania2019complex}.

In the $H/L\!\to\!\infty$ limit, a 2D frictional system features a single geometric scale --- the length $L$ --- and energy radiated by frictional rupture away from the interface propagates (through elastic waves) effectively indefinitely, without being reflected back and potentially affecting the interfacial dynamics. Consequently, one may wonder whether having a finite height $H$ would make a qualitative difference in relation to the emergence of complexity. Indeed, it was shown in~\cite{horowitz1989slip} that strictly homogeneous 2D frictional systems of the type discussed above feature long-term complex, non-periodic solutions in the quasi-static limit for $H/L\!\ll\!1$. The quasi-static limit neglects material inertia, which is equivalent to considering diverging elastic wave-speeds (in particular the shear wave-speed, $c_{\rm s}\!\to\!\infty$).

The geometry of $H/L\!\ll\!1$ systems corresponds to a long strip configuration, which is relevant for many engineering systems and invoked in the context of various geophysical problems/models (sometimes termed ``elastic slab models'' and ``elastic crustal plane models'')~\cite{shaw2000existence,rice1980mechanics,horowitz1989slip,lehner1981stress,Rice1983,johnson1992influence}. It is specifically relevant for natural faults that host elongated earthquake ruptures, such as subduction zone megathrust, long antiplane dip‐slip and strike‐slip faults, where the finite seismogenic width plays an analogous role to that of the strip's height $H$~\cite{Weng2017,Weng2019}.

The findings of~\cite{horowitz1989slip} were reinforced and extended in~\cite{shaw2000existence}, where the same 2D homogeneous frictional systems featuring $H/L\!\ll\!1$ have been studied using scalar elastodynamics with bulk dissipation/damping (instead of invoking the quasi-static approximation). It was concluded that non-periodic, irregular large slip bursts/events are generic (i.e., not specific to a narrow range of the frictional parameters) in continuum rate-and-state models for sufficiently small $H/L$.

We aim at studying the 2D homogeneous frictional system that gave rise to coarsening and regular long times solutions for $H/L\!\to\!\infty$, but in the opposite limit of $H/L\!\ll\!1$. In view of the results of~\cite{horowitz1989slip,shaw2000existence}, we expect that sufficiently reducing the height-to-length ratio $H/L$ would induce a coarsening-to-complexity transition. That is, we expect that for sufficiently small $H/L$, coarsening becomes less effective and the regular solution is dynamically avoided such that instead the system settles into a stochastic, statistically
stationary state. Our main goal is to gain deeper physical insight into the origin and nature of the coarsening-to-complexity transition and the emerging statistical properties.

We show that indeed sufficiently reducing $H/L$ leads to a coarsening-to-complexity transition, and in particular that for $H/L\!\sim\!{\cal O}(0.01\!-\!0.1)$ the 2D homogeneous system reveals stationary complexity. We find that the slip bursts can be classified into two types; small, predominantly non-propagating bursts that are power-law distributed and large, propagating bursts that are log-normally distributed. The small slip bursts/events feature a spatial pattern characteristic of rupture (crack-like slip function), even though they do not significantly increase their length during the event. They are responsible for a non-negligible fraction of the overall slip budget of the system and are shown to play dynamic roles in preparing the interface for nucleating large, propagating slip bursts/events.

We further demonstrate how spatiotemporal heterogeneity in the interfacial stress and state fields is self-generated, and how it affects frictional rupture nucleation and arrest. We show that the finite geometric scale $H$ systematically affects both the dynamics and the statistics of large events, as implied by fracture mechanics in the long strip configuration. We provide evidence that indicates that the effect of $H$ is not purely geometric, but also dynamic, through an inertial timescale. In particular, we show that reducing the elastic wave-speeds has a similar dynamical effect on the statistics as increasing $H$, indicating that the wave reflection (inertial) timescale $\sim\!H/c_{\rm s}$ that carries information regarding the finite geometry is of importance.

We discuss the robustness of our findings against variations of the frictional parameters, most notably in relation to the magnitude of frictional rate-weakening. We demonstrate that our results are semi-quantitatively similar for the two widely used interfacial state evolution laws within the rate-and-state friction framework, the so-called aging and slip laws. Finally, we demonstrate a reverse transition in which statistical complexity disappears in favor of the spatially periodic traveling solution, and discuss its possible physical origin. Taken together, these findings shed basic light on the emergence of complexity in simple, homogeneous frictional systems.
%%%%%%%%%%%%%%%%%%%%%%%% Figure %%%%%%%%%%%%%%%%%%%%%%%%%%%%%%%%%%
\begin{figure*}[ht!]
    %\noindent\makebox[0.5\textwidth]{%
    \centering
    \includegraphics[width=1\textwidth]{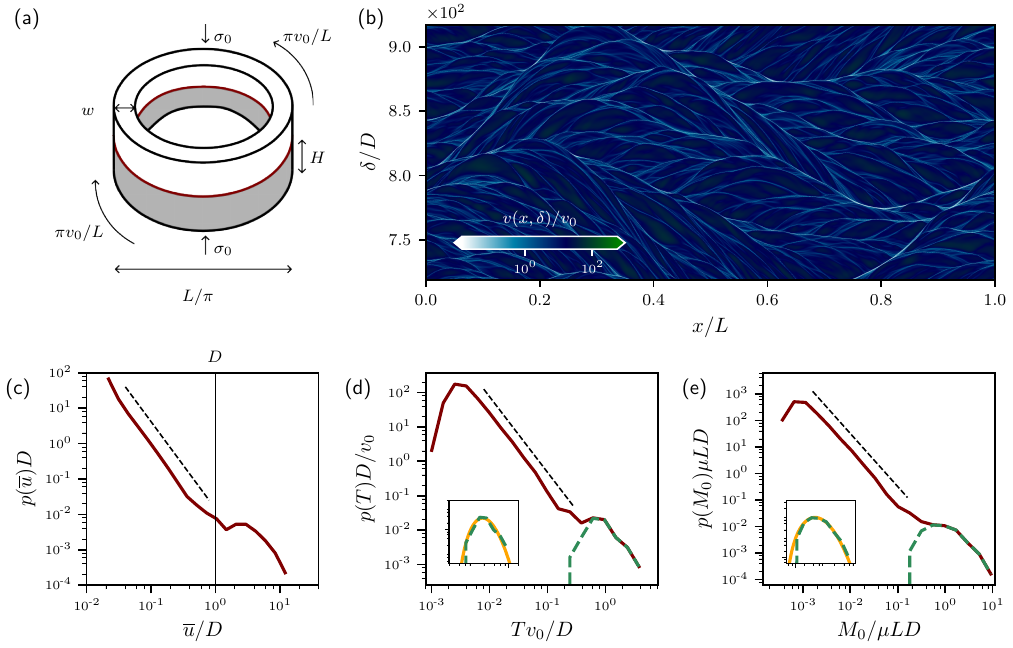}
    \caption{(a) A rotary (periodic) frictional system, composed of two identical deformable annuli (white and gray are just used to make a visual distinction) of diameter $L/\pi$, height $H$ and thickness $w$, is sketched. The two annuli are pressed one against the other by a normal stress $\sigma_0$ and each is driven by an angular velocity of magnitude $\pi v_0/L$ (anti-symmetrically). The frictional interface is marked in brown. The problem we solved corresponds to the 2D limit, i.e., $w\!\ll\!H,L$. (b) A space-slip plot of the slip velocity $v(x,\delta)/v_0$ in the $x/L\,$--$\,\delta/D$ plane ($x$ is the coordinate along the interface and $D$ is the characteristic slip distance). See color code, text and~\ref{sec:SM_identification}) for additional details. (c) The stationary probability distribution function $p(\bar{u})$ of the average slip $\bar{u}$ of slip events in our system with $H\!=\!0.1$ m, $L\!=\!4$ m and $v_0\!=\!2\times\!10$\textsuperscript{-3} m/s. The axes (note the double-logarithmic scale) are made dimensionless as indicated in the figure and the vertical solid line corresponds to $\bar{u}\!=\!D$. The dashed line, added as a guide to the eye, corresponds to a power-law with an exponent of $\simeq\!-2.45$. (d) As in panel (c), but for the duration $T$ of slip events. The dashed line corresponds to a power-law with an exponent of $\simeq\!-2.2$. The corresponding distribution of slip events featuring $\bar{u}\!>\!D$ is superimposed (dashed green line, see text for details). The inset presents a log-normal fit (solid yellow line) to the latter. (e) As in panel (d), but for the seismic moment $M_0$ of slip events. The dashed line corresponds to a power-law with an exponent of $\simeq\!-2$.}
\label{fig:fig1}
\end{figure*}
%%%%%%%%%%%%%%%%%%%%%%%%%%%%%%%%%%%%%%%%%%%%%%%%%%%%%%%%%

\vspace{0.25cm}
%\section*{Results}
\hspace{-0.35cm}{\bf \Large Results}

We consider two identical, homogeneous, 2D linear elastic bodies in frictional contact under the application of a compressive stress of magnitude $\sigma_0$. The frictional interface is located at $y\=0$ and extends along $x$, and each body features height $H$ (in the $y$ direction) and length $L$ (in the $x$ direction). The bodies are driven anti-symmetrically at $y\=\pm H$ by a tangential velocity $\pm v_0/2$. Finally, periodic boundary conditions along $x$ are employed.

Our system is analogous to a rotary (periodic) frictional system, composed of two identical linear elastic annuli of diameter $L/\pi$, height $H$ and thickness $w$, and driven anti-symmetrically by an angular velocity of magnitude $\pi v_0/L$. The rotary frictional system, which can be realized in the laboratory, is sketched in Fig.~\ref{fig:fig1}a, and formally maps to our system in the 2D limit corresponding to $w\!\ll\!H,L$. The vectorial linear elastodynamic problem in the two bodies, resulting in a 2D displacement vector field ${\bm u}(x,y,t)$, is solved using the explicit dynamic finite element method (FEM) framework, based on the in-house, open-source FEM library ``Akantu''~\cite{Richart2015,Rezakhani2020} (see Sect.~\ref{sec:SM_fem}) for additional details), once the frictional/interfacial boundary conditions are specified.

The two bodies are coupled at the interface, which is described at the continuum level through a rate-and-state dependent frictional shear resistance/strength $\tau(v,\phi)$. Here $v(x,t)\=\partial_t\delta(x,t)$ is the slip velocity field, where the slip field is $\delta(x,t)\!\equiv\!u_x(x,y\!\to\!0^+,t)-u_x(x,y\!\to\!0^-,t)$ (the superscripts $+/-$ correspond to the upper/lower bodies, respectively). $\phi(x,t)$ is an interfacial state field representing the contact age (of time dimensions), whose logarithm is related to the real contact area~\cite{Dieterich1994,Marone1998a,Baumberger2006,Bowden2001,Ben-david2010} and that follows the ``aging evolution law''~\cite{Dieterich1986,Marone1998a,Baumberger2006}
\begin{equation}
\partial_t\phi(x,t)=1-\frac{\phi(x,t)\,v(x,t)}{D} \ .
\label{eq:aging_law}
\end{equation}
The latter (see also~\ref{sec:SM_friction_law}) introduces a characteristic slip distance $D$ for the evolution of the interfacial state. The precise expression for $\tau(v,\phi)$ is given in Sect.~\ref{sec:SM_friction_law}, but the important point to note here is that under steady sliding $\tau$ is rate-weakening, $d\tau_{\rm ss}/dv\!<\!0$, over a broad range of slip velocities, including $v_0$. This implies that the frictional system is linearly unstable against infinitesimal perturbations with a wavelength larger than $L_{\rm c}(H)\!\ll\!L$~\cite{Aldam2017}, which is a necessary condition for the emergence of complexity.

In our calculations, the initial slip velocity and state fields are set to their homogeneous steady-state values, i.e., $v(x,t\=0)\=v_0$ and $\phi(x,t\=0)\=D/v_0$, respectively. We add to $\phi(x,t\=0)$ a small amplitude, Gaussian white noise in space. The latter is not necessary, i.e., the linear elasto-frictional instability would be triggered also by numerical noise, but it speeds up transient dynamics. After a finite time transient, the system settles into a stochastic, statistically-stationary state, which is analyzed next.

\vspace{0.3cm}
%\subsection*{Statistical complexity}
\hspace{-0.35cm}{\bf Statistical complexity}
\vspace{0.3cm}

To first offer some visual representation of the emerging statistical complexity in our problem and to introduce our analysis procedure, we present in Fig.~\ref{fig:fig1}b a space-slip plot corresponding to one of our numerical simulations with $H\!=\!0.1$ m and $L\!=\!4$ m, satisfying $H/L\!\ll\!1$, and $v_0\=2\times10$\textsuperscript{-3} m/s. To start quantifying the complex dynamics, we compute at each point $x$ along the interface at any time $t$ the time interval it takes the interface to reach a minimal slip increment. The ratio between the two allows us to assign a slip velocity $v(x,\delta)$ at a given accumulated slip $\delta$. We then generate a color map of $v(x,\delta)/v_0$ in the space-slip $x$-$\delta$ plane. The dark blue and green end of the color code corresponds to rapid slip, much larger than the applied velocity $v(x,\delta)/v_0\!\gg\!1$, and dark blue/green patches in the space-slip plot correspond to slip occurring roughly at the same time. On the other hand, the bright/white end of the color code essentially corresponds to a sticking state, $v(x,\delta)/v_0\!\ll\!1$, hence the bright white lines indicate non-sliding interfacial patches.

While Fig.~\ref{fig:fig1}b offers some visual evidence for the complex and heterogenous slip dynamics in the system, it still requires additional analysis in order to make things quantitative. To that aim, we use the extracted time intervals information to define discrete slip bursts/events (we will use ``bursts'' and ``events'' interchangeably hereafter), as explained in detail in Sect.~\ref{sec:SM_identification}. We verified that the actual values of the thresholds/increments invoked in the operational definition of slip events have no qualitative effect on the obtained results, see Sect.~\ref{sec:SM_threshold_effect}. For each slip event, we compute the spatial averaged slip $\bar{u}$, the time duration $T$ and the spatial extent $L_{\rm s}$ (`s' stands for `slip'). The seismic moment of each slip event is computed as $M_0\!\equiv\!\mu L_{\rm s} \bar{u}$. The probability distribution functions $p(\bar{u})$, $p(T)$ and $p(M_0)$ are presented in Fig.~\ref{fig:fig1}c-e, respectively. All quantities are nondimensionalized by the corresponding natural parameters, as indicated by the axis labels. The statistical distributions are time-translational invariant in the long-time limit, see Sect.~\ref{sec:SM_stationarity}.
%%%%%%%%%%%%%%%%%%%%%%%%% Figure %%%%%%%%%%%%%%%%%%%%%%%%%
\begin{figure*}[ht!]
    %\noindent\makebox[\textwidth]
    {%
    \centering \includegraphics[width=\textwidth]{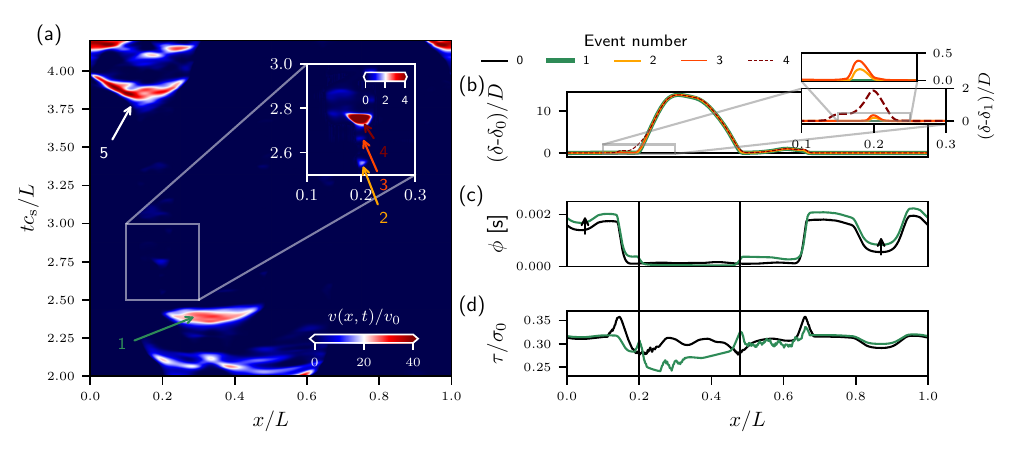}}
    \caption{(a) A space-time plot of $v(x,t)/v_0$ in the $x/L\,$--$\,t c_{\rm s}/L$ plane (corresponding to a fraction of a simulation, see $y$-axis range) for a system with $H\!=\!0.1$ m and $L\!=\!2$ m. A zoom in on a small portion of the dynamics is presented in the inset, also involving a change in the scale of the color code, as indicated by the two color bars. Five slip events are marked, see extensive discussion in the text. (b) The accumulated slip $\delta(x)/D$ associated with the first four slip events (see upper legend for the line types and colors), relative to previously accumulated slip ($\delta_0(x)$ and $\delta_1(x)$ are defined in the text). The two insets in the upper corner present zoom in views as indicated by the gray boxes and lines, see extensive discussion in the text. (c) The internal state field $\phi(x)$, before (black) and after (green) event 1. The arrows indicate contact aging. The two vertical black lines mark the locations of the rupture edges associated with event 1 upon arrest. See extensive discussion in the text. (d) The same as panel (c), but for $\tau(x)/\sigma_0$. See also Supplemental Movie1, 2 and 3 in Sect.~\ref{sec:movies}, in relation to events 1, 4 and 5 marked in panel (a).}
\label{fig:fig2}
\end{figure*}
%%%%%%%%%%%%%%%%%%%%%%%%%%%%%%%%%%%%%%%%%%%%%%%%%%%%%%%%%%%%%%

The distributions share a few generic properties. First, for small values of each dimensionless observable, the distribution follows a power-law, as indicated by the dashed lines that are added as guides to the eye (note the double-logarithmic scale used and see figure caption for additional details). Second, at larger values of each dimensionless observable, the distribution features a rather broad ``hump'' that appears to be distinguished from the power-law at small values. Consequently, our first goal is to understand the physical origin and nature of the distinction between the two parts of each distribution.

To that aim, we note that the two parts of $p(\bar{u})$ appear to be approximately distinguished by being smaller or larger than $\bar{u}\=D$ (marked by the vertical solid line in Fig.~\ref{fig:fig1}c). During propagating frictional rupture, one expects the typical slip displacement at a point along the interface to be significantly larger than the characteristic slip $D$, and to be accompanied by rather significant strength reduction. Consequently, we expect slip events that feature $\bar{u}\!<\!D$ --- which are predominantly power-law distributed --- to correspond to non-propagating frictional rupture.

That is, while these slip events feature a maximal slip rate larger than the applied velocity $v_0$, their spatial extent $L_{\rm s}$ does not significantly increase over their duration $T$. A corollary of this expectation is that the ratio $L_{\rm s}/T$ associated with these slip events is not constrained by causality --- i.e., by elastic wave-speeds --- in contrast to propagating frictional rupture. Indeed, we found that $L_{\rm s}/T$ that is associated with slip events with $\bar{u}\!<\!D$ attains values that are significantly larger than the dilatational wave-speed $c_{\rm d}$ (not shown).

The latter observation clearly supports the predominantly non-propagating nature of slip events with $\bar{u}\!<\!D$. Consequently, we use the latter as an operational definition of small, non-propagating slip events. The complementary regime, $\bar{u}\!>\!D$, approximately corresponds to larger, propagating slip events. In order to use this operational definition to quantify things, we screen small slip events according to $\bar{u}\!<\!D$, and superpose the $p(T)$ and $p(M_0)$ distributions for the remaining slip events in Fig.~\ref{fig:fig1}d-e, appearing as dashed green lines. The latter are predominantly parabolic (recall the double-logarithmic scale used) and hence indicate a log-normal distribution, as demonstrated in the corresponding insets (see figure caption for details). The log-normal distribution of large slip events appears to be consistent with the findings of~\cite{shaw2000existence}, cf.~Fig.~2 therein.

The above analysis, summarized in Fig.~\ref{fig:fig1}c-e, provides a comprehensive picture of the statistics of slip events in our problem. It shows that the slip events can be approximately classified into small, predominantly non-propagating events, whose properties follow power-law distributions and larger, predominantly propagating events, whose properties follow log-normal distributions (i.e., the logarithm of an observable is normally distributed). This nontrivial, statistical steady state (time-translational invariant) stands in sharp and qualitative contrast to the corresponding behavior of the very same system in the $H\!\to\!\infty$ limit. As stated above, in the latter case the system undergoes continuous coarsening until settling into a deterministic steady state. The latter features a single slip pulse steadily propagating through the periodic boundary conditions~\cite{Roch2022}, implying that all relevant observables are essentially $\delta$-function distributed. Our next goal is to gain physical insight into this dramatic difference, following~\cite{horowitz1989slip,shaw2000existence}, and hence into the emergence of some aspects of complexity in the finite $H$ system.

\vspace{0.3cm}
%\subsection*{The spontaneous self-generation of heterogeneity and dynamical complexity}
\hspace{-0.35cm}{\bf The spontaneous self-generation of heterogeneity and dynamical complexity}
\vspace{0.3cm}

The problem under consideration is strictly spatially homogeneous, i.e., neither the material and interfacial properties, nor the geometry and loading conditions feature any heterogeneity/disorder. Consequently, the challenge is to understand how heterogeneity/disorder is dynamically self-generated in the finite $H$ system, while it is absent in the $H/L\!\to\!\infty$ case (in the long-time limit). Moreover, one needs to understand the physical mechanisms that tend to arrest frictional rupture in our problem, preventing it from being system spanning.

To address some of these issues, we first present in Fig.~\ref{fig:fig2}a a space-time plot of the slip velocity field $v(x,t)/v_0$ in the $x\,$--$\,t$ plane. We focus on a small time interval ($t c_{\rm s}/L\=2.00-4.25$ therein) in order to isolate the spatiotemporal dynamics of a few slip events. We first focus on the slip event marked as ``event 1'' (in green) in Fig.~\ref{fig:fig2}a. This event, which was accompanied by a maximal slip velocity of more than an order of magnitude larger than the applied one (see color bar), nucleated at $x/L\!\simeq\!0.35$, and propagated bi-laterally until being arrested at $x/L\!\simeq\!0.2$ (left edge) and slightly before $x/L\!\simeq\!0.5$ (right edge), both marked by vertical black lines in Fig.~\ref{fig:fig2}c-d (to be discussed below, see also Supplemental Movie1 in Sect.~\ref{sec:movies}).

In Fig.~\ref{fig:fig2}b, we plot the slip accumulated by this propagating slip event (thick green line), relative to the slip $\delta_0(x)$ accumulated in the history prior to this event (it is represented by the horizontal straight black line in Fig.~\ref{fig:fig2}b, even though it is spatially varying in itself, and the entire history is denoted as ``event 0'' in the upper legend). The slip accumulated by event 1 is approximately proportional to the functional form $\sqrt{(L_{\rm s}/2-\Delta{x})(L_{\rm s}/2+\Delta{x})}$, where $L_{\rm s}$ is the spatial extent of the slip event and $\Delta{x}$ is measured from center (close to the nucleation location). This functional form (the quantitative fit itself is not shown) is a clear signature of conventional crack-like scaling~\cite{Broberg1999}.

The slip event under discussion (event 1), did not propagate into a spatially homogeneous state, i.e., the complex spatiotemporal dynamics prior to it generated nontrivial interfacial fields $\phi(x)$ and $\tau(x)$, which are plotted in black in Fig.~\ref{fig:fig2}c-d, respectively. We do not discuss here how these fields were generated by prior dynamics (looking at the time interval $t c_{\rm s}/L\=2.00-2.25$ in Fig.~\ref{fig:fig2}a provides a clear hint, though), but rather stress that they feature self-generated spatial heterogeneity in a strictly homogeneous system (in terms of the material parameters and geometry) and serve as initial conditions for event 1, carrying memory of past dynamics. The resulting $\phi(x)$ and $\tau(x)$ immediately after rupture arrest --- the arrest locations are marked by the vertical black lines, as noted above --- are superimposed in green.

The arrest locations appear to coincide with local minima of the stress field $\tau(x)/\sigma_0$ (black curve in Fig.~\ref{fig:fig2}d) into which event 1 propagated, possibly indicating a causal relation between the two, though we cannot exclude that other physical processes were at play. In fact, the duration of event 1 is comparable with the back-and-forth elastic shear waves travel time $2H/c_{\rm s}$, an important point to be further elaborated on below. Outside the slipping domain of extent $L_{\rm s}$, the interface has experienced contact aging under nearly stick conditions (increase in the contact time $\phi$), as indicated by the upward arrows in Fig.~\ref{fig:fig2}c. Finally, as expected from a crack-like object, the arrested event 1 gave rise to significant stress concentration near the arrest locations (see local maxima of the green curve in Fig.~\ref{fig:fig2}d at the vertical black lines).

Looking at Fig.~\ref{fig:fig2}a, one may get the impression that the interface was quiescent during the time interval $t c_{\rm s}/L\=2.50-3.75$, after event 1 was arrested and before another large, propagating slip event nucleated (marked as ``event 5''). This is, however, not the case. In fact, quite a few smaller slip events took place during this time interval, they are just not clearly visible on the scale of $v(x,t)/v_0\!\simeq\!40$ (see main color bar) that characterizes large events. Next, we focus on slip events that took place near the left arrest location of event 1, $x/L\!\simeq\!0.2$, which are likely to be affected by the stress concentration that remained there by event 1.

We zoom in on this spatial location over the time interval $t c_{\rm s}/L\=2.50-3.00$, as shown in the inset in the upper right corner of Fig.~\ref{fig:fig2}a. Note that due to the broad span of slip event properties, we changed the color bar to correspond to slip velocities up to $v(x,t)/v_0\=4$, which is an order of magnitude smaller than the range in the main panel. We marked in the inset three consecutive slip events, denoted as events 2, 3 and 4 therein (see also the events legend above Fig.~\ref{fig:fig2}b). The resulting slip field after each of these events is superimposed on Fig.~\ref{fig:fig2}b (see line types and colors in the upper legend). To get a better picture of the dynamics, we added a zoom in inset (upper right corner), where we focus on $x/L\=0.1-0.3$ and plot the slip accumulated by these events relative to $\delta_1(x)$, left after event 1 (consequently, this time, the history is represented by the horizontal straight green line).

The maximal slip accumulated by events 2 and 3 is small, hence we further zoom in on the field in the second inset. The latter suggests that events 2 and 3 are non-propagating and accommodate slip less than $D$ (note the $y$-axis of the uppermost inset). Yet, their slip also approximately follows a crack-like scaling proportional to $\sqrt{(L_{\rm s}/2-\Delta{x})(L_{\rm s}/2+\Delta{x})}$, indicating that they are in fact crack-like objects of an approximately fixed spatial extent $L_{\rm s}$ (compare the slip field of the propagating event 4 to that of the non-propagating event 3 that preceded it).

Event 4 did involve propagation away from the arrest location of event 1 at $x/L\!\simeq\!0.2$, and was accompanied by a significantly larger slip velocity (see inset in Fig.~\ref{fig:fig2}a and Supplementary Movie2) and accumulated slip (with a peak value of $2D$). The nucleation dynamics of slip events 2, 3 and 4 appear to differ from the homogeneous nucleation scenario controlled by the elasto-frictional length $L_{\rm c}$. Instead, their nucleation seems to be influenced by large stress gradients accompanying static stress concentrations left by previous events. This is most pronounced in relation to event 4, see Supplementary Movie2. Better understanding nucleation dynamics in our system is a challenge for future work. Clarifying the role of dynamical noise as a triggering mechanism, related to elastic waves that propagate within the system (to be further discussed below), is yet another topic for future studies.

The accumulated effect of the relatively small slip events plays a role in preparing the interface for subsequent larger, propagating slip events. In the example in Fig.~\ref{fig:fig2}a, they are likely to give rise to the next large, propagating slip event (event 5, to be further discussed below). Their broad statistical distribution may reflect the properties of the emergent dynamical noise in the system, possibly related to chaotic motion of small-amplitude elastic waves. Moreover, all of the small events in this simulation, i.e., those featuring $\bar{u}\!<\!D$ in Fig.~\ref{fig:fig1}c, accounted for $8\%$ of the total slip accommodated by the interface over the entire simulation. The small, non-propagating events were not previously discussed in~\cite{shaw2000existence}, where they were apparently excluded by choosing large thresholds in operationally defining slip events. This choice may reflect the focus of~\cite{shaw2000existence} on slip events with clear seismic signatures. Moreover, dynamical noise in~\cite{shaw2000existence} was minimized by including bulk dissipation/damping.

Overall, the complex spatiotemporal dynamics discussed in relation to Fig.~\ref{fig:fig2} demonstrate a hierarchy of slip events, featuring a broad range of properties, which interact among themselves to give rise to the statistical complexity presented in Fig.~\ref{fig:fig1}. Moreover, they demonstrate the emergence of spontaneously self-generated heterogeneity/disorder in a system that features no quenched disorder or bulk nonlinearity. The fluctuations associated with self-generated heterogeneity appear to contribute to complexity in both triggering slip events and arresting them. Since self-generated heterogeneity/disorder is absent for $H/L\!\to\!\infty$ over long times, where coarsening is efficient and a regular solution is obtained, its origin must be related to the finite height $H$ in our system. Consequently, we next discuss the roles played by the geometric length $H$ in affecting the magnitude of slip events and in rupture arrest.

\vspace{0.3cm}
%\subsection*{Elastic wave reflections and the effect of\\ the long strip geometry}
\hspace{-0.35cm}{\bf The long strip geometry and the elastic wave reflections}
\vspace{0.3cm}

As noted earlier, the $H/L\!\ll\!1$ limit corresponds to a long strip configuration, which is relevant for many engineering and geophysical problems (e.g., when elongated earthquake ruptures saturate at the fault's seismogenic width~\cite{Weng2017,Weng2019}). This configuration was extensively studied in a related context, that of classical fracture mechanics~\cite{freund1998dynamic,Broberg1999}, and hence it may be useful to explore its possible implications for our problem. In the classical fracture mechanics problem, rupture dynamics are controlled by the elastic energy density per unit length stored in the strip, which is proportional to the strip height $H$ under fixed-grip boundary condition~\cite{freund1998dynamic,Broberg1999}. This boundary condition is relevant to our problem since the propagation velocity of large slip events and their characteristic slip velocity are large compared to the loading velocity $v_0$, implying that during their lifetime the strip is loaded by an approximately constant displacement.

In the fracture mechanics problem, when rupture is smaller than $H$, the system is effectively infinite from the rupture perspective and the stress concentration near the rupture edges is controlled by its length. In the opposite limit, when rupture is much larger than $H$, active deformation takes place on a scale $H$ near the rupture edges (similarly to a pulse of size $\sim\!H$) and the stress concentration is controlled by $H$. That is, in this case rupture is controlled by the finite geometric scale $H$, which also affects whether propagation takes place or not, depending on the available energy density~\cite{freund1998dynamic,Broberg1999}.

The information regarding the finite strip geometry and the finite amount of energy density available is carried by elastic waves that are radiated from the nucleating rupture and bounce back from the strip boundaries at $y\=\pm H$ (instead of being radiated away from the interface indefinitely in the $H/L\!\to\!\infty$ limit). The time interval associated with this process corresponds to the travel time of reflected shear waves, i.e., $2H/c_{\rm s}$. If these fracture mechanics considerations are relevant for our frictional problem, then the timescale $2H/c_{\rm s}$ should have a clear signature in the dynamics of the system and the emerging statistics.

A first hint that this is indeed the case was already provided by event 1, whose duration is comparable to $2H/c_{\rm s}$. Moreover, the dynamics of event 5 in Fig.~\ref{fig:fig2}a (see also Supplementary Movie3) also support this connection. This rupture event propagated bi-laterally for a certain amount of time until it transformed into two propagating slip pulses of size comparable to $H$ (the size roughly corresponds to the horizontal width of each of the red patches away from the center of the event, and recall that here $H\=0.1$ m and $L\=2$ m). The duration of the bi-lateral propagation stage, roughly corresponding to the vertical width of the red patch at the center of event 5, is comparable to $2H/c_{\rm s}$, which corresponds to ${\cal O}(0.1)$ in units of $tc_{\rm s}/L$.

Event 5 was spontaneously nucleated inside a spatially heterogeneous state, self-generated by the slip history that preceded it, and in the presence of dynamical background noise. It would be desired to first test the relevance of the fracture mechanics strip considerations in a more controlled manner. To this aim, we performed calculations on the very same frictional system, but with a deterministic initial perturbation centered at $x/L\=0.5$. The latter features a spatial extent larger than $L_{\rm c}(H)$ such that an individual propagating slip event (frictional rupture) is nucleated in a controlled manner. Moreover, we focus on the subsequent dynamics over relatively short times, before significant heterogeneity and noise are built up. As such, we isolate the effect of the interaction of a propagating slip event with elastic waves that it radiates itself and are back reflected from finite boundaries.
%%%%%%%%%%%%%%%%%%%%%%%%% Figure %%%%%%%%%%%%%%%%%%%%%%%%%
\begin{figure*}[ht!]
    %\noindent\makebox[\textwidth]{%
    \centering \includegraphics[width=1\textwidth]{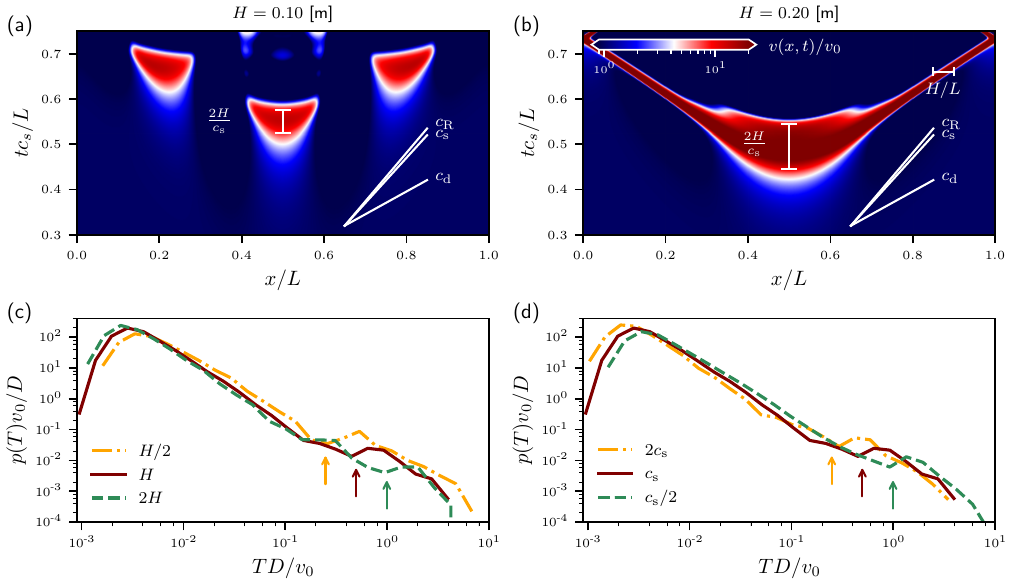}
    \caption{(a) A space-time plot of $v(x,t)/v_0$ (same format as in Fig.~\ref{fig:fig1}a) for a system of height $H\!=\!0.1$ m and $H\!=\!4$ m, where a deterministic $t\!=\!0$ perturbation is introduced around $x/L\!=\!0.5$. The color bar is the same as the one appearing in panel b. The wave reflection time $2H/c_{\rm s}$ (in the units used in the $y$-axis) is marked by the vertical white bar at the middle of the first slip event (two additional slip events occurring at a later time around $x/L\!=\!0.2$ and $x/L\!=\!0.8$ are not discussed). The 3 elastic wave-speeds: $c_{\rm R}$ (Rayleigh), $c_{\rm s}$ (shear) and $c_{\rm d}$ (dilatational) are marked by the inclined white lines. See extensive discussion in the text. (b) The same as panel a, but for $H\!=\!0.2$ m, see extensive discussion in the text. A horizontal white bar of size $H$ is added on the right. See extensive discussion in the text. (c) $p(T)$ for 3 values of $H$ as indicated in the legend, where the reference case (solid brown line) is identical to Fig.~\ref{fig:fig1}d (where $H\!=\!0.1$ m). The upward pointing arrows mark the values of $2H/c_{\rm s}$ (same color code as in the legend). See text for discussion. (d) The same as panel (c), but for 3 values of $c_{\rm s}$ as indicated in the legend, where the upward pointing arrows mark the values of $2H/c_{\rm s}$. See text for discussion.}
    \label{fig:fig3}
\end{figure*}
%%%%%%%%%%%%%%%%%%%%%%%%%%%%%%%%%%%%%%%%%%%%%%%%%%%%%%%%%

The results are presented in Fig.~\ref{fig:fig3}a, where a space-time plot of $v(x,t)/v_0$ of the emerging dynamics with $H\=0.1$ m is shown (same format as in Fig.~\ref{fig:fig2}a). The slip velocity significantly increases around $x/L\=0.5$ (the color bar is the same as in panel (b) and note the logarithmic scale used) and starts to rapidly propagate bi-laterally (the elastic wave-speeds are marked, see figure caption), as indicated by the expanding red patch in the figure. Yet, after a certain amount of time (roughly corresponding to the size of the red patch along the $y$-axis) frictional rupture arrests. The duration of the slip event, from the time the slip velocity becomes sizable until arrest is comparable to the wave reflection time $2H/c_{\rm s}$, which is marked by the vertical white bar in the middle of the slip event. That is, similarly to events 1 and 5 in Fig.~\ref{fig:fig2}a, this slip event reveals a timescale that is consistent with $2H/c_{\rm s}$.

At the same time, and differently from event 5 in Fig.~\ref{fig:fig2}a, the slip event arrests before transforming into two propagating slip pulses (two other slip events, which are not discussed here, take place later around $x/L\!\simeq\!0.2$ and $x/L\!\simeq\!0.8$). This indicates that when reflected waves reached the nucleating slip event, the amount of available energy was not large enough to support propagation (recall that the system is continuously loaded at a velocity $v_0$). Since the latter can be increased by increasing $H$, we repeated the calculation discussed in Fig.~\ref{fig:fig3}a with everything being identical, except for doubling $H$ (i.e., using $H\=0.2$ m).

The resulting dynamics are presented in Fig.~\ref{fig:fig3}b. It is observed that slip at the middle of the event stops after a time interval comparable to the wave reflection time $2H/c_{\rm s}$ (marked again by the vertical white bar), but this time two pulses subsequently propagate across the system (under the employed nucleation conditions and available energy, and in the absence of self-generated heterogeneity, they traverse the entire interface). The pulses are supershear, i.e., propagate at a velocity close to the dilatational wave-speed $c_{\rm d}$ (see figure and its caption), and feature a characteristic size $H$ (as indicated by the horizontal white bar added on top of the right propagating pulse).

Taken together, the results in Fig.~\ref{fig:fig3}a-b are in agreement with the long strip fracture mechanics considerations, highlighting the importance of the finite geometric scale $H$ and the dynamical/inertial timescale $\sim\!H/c_{\rm s}$. Moreover, our findings appear to be mirrored in the context of strike-slip faults, where it has been shown~\cite{Weng2017} that relatively narrow seismogenic zones (analogous to small $H$) lead to self-arresting ruptures, as in Fig.~\ref{fig:fig3}a, and that wider seismogenic zones (analogous to larger $H$) lead to breakaway ruptures, i.e., to pulses that propagate over long distances and inherit their scale from the finite seismogenic width, as in Fig.~\ref{fig:fig3}b.

The above-discussed physical picture suggests that the timescale $\sim\!H/c_{\rm s}$ would manifest itself also in the statistics of large slip events. To test this expectation, we plot in Fig.~\ref{fig:fig3}c the event duration probability distribution $p(T)$ for 3 values of $H$. The original data previously presented in Fig.~\ref{fig:fig1}d appear in the solid brown line and are marked as `$H$' in the legend. The two other distributions correspond to halving and doubling this reference value of $H$ (everything else being fixed, including $c_{\rm s}$), see legend. It is observed that the peak of the large, propagating slip events part of $p(T)$ is systematically shifted to higher values with increasing $H$ (the values corresponding to $2H/c_{\rm s}$ are marked by the colored arrows, see figure caption).

If indeed the wave reflection time $2H/c_{\rm s}$ is of basic importance, then reducing/increasing $c_{\rm s}$ at a fixed $H$ by the same factor that $H$ is increased/reduced at a fixed $c_{\rm s}$, respectively, should have a similar effect on $p(T)$. To test this, we plot $p(T)$ in Fig.~\ref{fig:fig3}d for 3 values of $c_{\rm s}$, this time the reference case is marked as `$c_{\rm s}$' in the legend. It is observed that doubling/halving $c_{\rm s}$ has a similar (though not strictly identical) effect on the large, propagating slip events part of $p(T)$ as that of halving/doubling $H$, respectively. A similar signature of the timescale $\sim\!H/c_{\rm s}$ in $p(\bar{u})$ is demonstrated in Sect.~\ref{sec:SM_reflection}. These results support the idea that the finite height $H$ limits the duration/size of large slip events, making coarsening less effective, and that the wave reflection time $2H/c_{\rm s}$ affects the dynamics and statistics of the emerging complex steady state. A finite $H$ also reduces the stress concentration near rupture edges~\cite{shaw2000existence}, potentially making dynamically-generated heterogeneity more effective in stopping them, as discussed in relation to Fig.~\ref{fig:fig2}.

\vspace{0.3cm}
%\subsection{On the robustness of complexity and a reverse complexity-to-coarsening transition}
\hspace{-0.35cm}{\bf On the robustness of complexity and a reverse complexity-to-coarsening transition}
\vspace{0.3cm}

The results presented above regarding the emergence of complexity in finite frictional systems in the absence of quenched disorder, and the physical understanding of its origin, suggest that it is might be weakly dependent on the details of the interfacial constitutive relation (friction law) if it satisfies the stated conditions. That is, our findings indicate that as long as the friction law retains its salient physical properties, most notably sufficiently strong rate-weakening and the incorporation of an interfacial state field $\phi(x,t)$ whose evolution is controlled by a finite slip distance $D$, complexity would emerge for small height-to-length ratios $H/L$.

To test this expectation, we first fixed the geometry and loading conditions as in Fig.~\ref{fig:fig1}, as well as the functional form of the frictional strength $\tau(v,\phi)$ (see Sect.~\ref{sec:SM_friction_law}) and the evolution of $\phi(x,t)$ according to Eq.~\eqref{eq:aging_law}. We then reduced the slip distance $D$ by a factor of 3 and varied the parameters of the frictional strength $\tau(v,\phi)$ such that friction becomes more rate-weakening (i.e., the total derivative of $\tau(v,\phi)$ with respect to $v$ under steady sliding conditions becomes more negative). The results are reported in Sect.~\ref{sec:varying_parameters} and demonstrate that complexity still emerges under these parameter variations. Moreover, the emerging statistical distributions are quantitatively similar to their counterparts in Fig.~\ref{fig:fig1} (see Sect.~\ref{sec:varying_parameters}, and Figs.~\ref{fig:figS10}-\ref{fig:figS12} in particular).
%%%%%%%%%%%%%%%%%%%%%%%%%%%%%%%%%%%%%%%%%%%%%%%%%%%%%%%%%%%%%%%%%%%%%%%%%%%%%%%%%
\begin{figure*}[ht!]
  \centering
    \includegraphics[width=\textwidth]{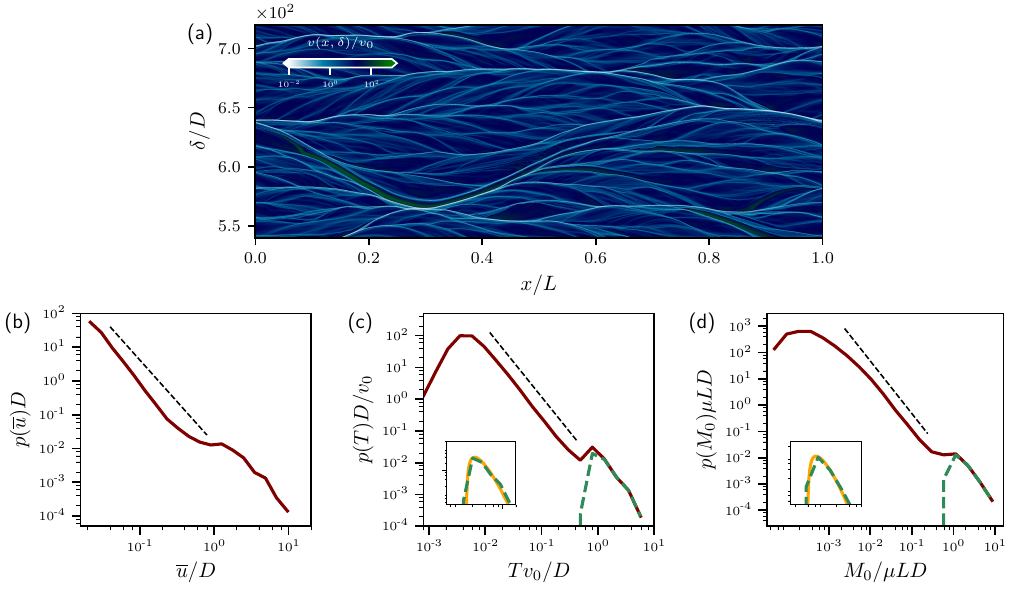}
  \caption{Results of a simulation identical to the one corresponding to Fig.~\ref{fig:fig1}, except that the slip evolution law of Eq.~\eqref{eq:slip_law} is used instead of the aging law of Eq.~\eqref{eq:aging_law}. (a) A space-slip plot of the slip velocity $v(x,\delta)/v_0$ in the $x/L\,$--$\,\delta/D$ plane, qualitatively demonstrating the emergence of complexity in the system, similarly to Fig.~\ref{fig:fig1}b. The plot reveals slip events that span a larger portion of the interface compared to those observed in Fig.~\ref{fig:fig1}b, see text for discussion. (b) The stationary probability distribution function $p(\bar{u})$. (c) The stationary probability distribution function $p(T)$. The corresponding distribution of slip events featuring $\bar{u}\!>\!D$ is superimposed (dashed green line), similarly to Fig.~\ref{fig:fig1}c. The inset presents a log-normal fit (solid yellow line) to the latter. (d) The stationary probability distribution function $p(M_0)$. The superimposed dashed lines in panels (b), (c) and (d) correspond to power-laws with the same exponents as in Fig.~\ref{fig:fig1}c-e to highlight the quantitative similarity between the two sets of results.}
  \label{fig:fig4}
\end{figure*}
%%%%%%%%%%%%%%%%%%%%%%%%%%%%%%%%%%%%%%%%%%%%%%%%%%%%%%%%%%%%%%%%%%%%%%%%%%%%%%%%%

Next, we test the robustness of our findings and physical picture against changes in the interfacial state evolution law, keeping the rate-weakening level and $D$ fixed. Specifically, we replace the aging evolution law of Eq.~\eqref{eq:aging_law} by the slip evolution law~\cite{Ruina1983}
\begin{equation}
\partial_t\phi(x,t) = -\frac{v(x,t)\,\phi(x,t)}{D}\log\!{\left(\frac{v(x,t)\,\phi(x,t)}{D}\right)}  \ ,
\label{eq:slip_law}
\end{equation}
while keeping $\tau(v,\phi)$ and the values of all friction parameters as in the reference case of Fig.~\ref{fig:fig1}. We also keep the geometry corresponding the the latter, i.e., $H\=0.1$ m and $L\=4$ m, fixed. The slip evolution law of Eq.~\eqref{eq:slip_law}, similarly to the aging law of Eq.~\eqref{eq:aging_law}, depends on the dimensionless combination $v(x,t)\,\phi(x,t)/D$ and specifically also incorporates the finite slip distance $D$. Yet, the functional form of the right-hand-side of Eq.~\eqref{eq:slip_law} is quite different. Most notably, it contains nonlinearity in the dimensionless combination $v(x,t)\,\phi(x,t)/D$, absent in Eq.~\eqref{eq:aging_law}, and features no restrengthening (aging) at stationary contact ($v\=0$).

The slip evolution law is known to give rise to quite different nucleation dynamics compared to the aging law (see, for example,~\cite{Ampuero2008}). Moreover, in view of the role of contact aging demonstrated in Fig.~\ref{fig:fig2}c, we expect the absence of aging (restrengthening under stationary conditions) to reduce the level of self-generated heterogeneity and hence to give rise to larger slip events. Yet, we expect this difference in the dynamics of the interfacial state field to lead to quantitative differences in the statistical distributions, but that the emerging complexity would retain its salient features. In Fig.~\ref{fig:fig4}, we present the results of a large-scale simulation exactly as the one corresponding to Fig.~\ref{fig:fig1} (including the values of the parameters stated in Table~\ref{tab:SM_parameters}, as mentioned above, and using the same presentational format as in Fig.\ref{fig:fig1}b-e), except that Eq.~\eqref{eq:slip_law} replaced Eq.~\eqref{eq:aging_law}. It is observed that while the slip evolution law gives rise to somewhat larger slip events as expected (see Fig.~\ref{fig:fig4}a), the statistical distributions in Fig.~\ref{fig:fig4}b-d are qualitatively --- and even semi-quantitatively (see figure caption) --- similar to their aging law counterparts of Fig.~\ref{fig:fig1}c-e. These results lend additional support to the generality of our findings and to the developed physical picture regarding the minimal conditions for the emergence of complexity.

\begin{figure*}[ht!]
  \centering
    \includegraphics[width=0.55\textwidth]{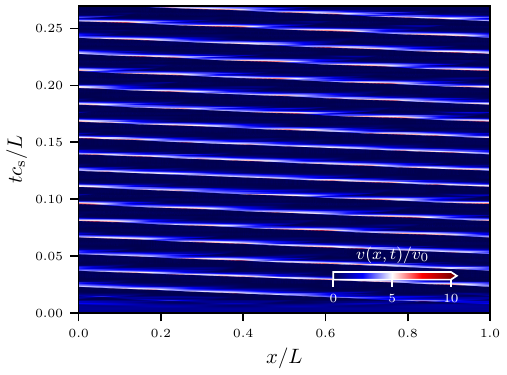}
  \caption{Space-time plot of $v(x,t)/v_0$ (same presentational format as in Fig.~\ref{fig:fig2}a and Fig.~\ref{fig:SM_03_ct_slipmap}a) for a simulation identical to the one corresponding to Fig.~\ref{fig:fig1}, except that we use $f_0\!=\!0.18$, $a\!=\!0.015$ and $b\!=\!0.116$, see Eq.~\eqref{eq:6_fem}. The latter are chosen such that their coordinated effect is to maintain $f_{\rm ss}(v_0)\!=\!\tau[v_0,\phi_{\rm ss}(v_0)]/\sigma_0$ at its reference value, but to {\em reduce} the rate-weakening magnitude (i.e., make the slope of the steady-state friction curve {\em less} negative). Specifically, the logarithmic rate-weakening was reduced from $df_{\rm ss}(v_0)/d\log\!{(v)}\!=\!-0.016$ (corresponding to Fig.~\ref{fig:fig1}, and in particular to $f_{\rm ss}(v)$ of Fig.~\ref{fig:SM_01_friction_law} and Table~\ref{tab:SM_parameters}) and $df_{\rm ss}(v_0)/d\log\!{(v)}\!=\!-0.00588$, i.e., by nearly a factor of 3. It is observed that under these conditions complexity disappears and instead the system settles, after a short coarsening transient, into a spatially periodic traveling pulse solution (that persistently propagates through the periodic boundary conditions).}
  \label{fig:fig5}
\end{figure*}

As discussed above, the minimal conditions for the emergence of complexity include sufficiently strong rate-weakening friction. While we did not quantify the precise rate-weakening threshold, which is surely $H$ dependent, it is clear that such a threshold exists (simply based on continuity, elasto-frictional instabilities and hence complexity are eliminated in the absence of rate-weakening friction). That is, we expect that by sufficiently reducing the magnitude of frictional rate-weakening at a fixed $H$ complexity would disappear, i.e., a reverse complexity-to-coarsening transition would be induced. To test this expectation, we present in Fig.~\ref{fig:fig5} the results of a simulation identical to the one corresponding to Fig.~\ref{fig:fig1}, except that the rate-weakening magnitude is quite significantly reduced, as indicated in the figure caption. It is observed that under these conditions complexity disappears and instead the system settles, after a short coarsening transient, into a spatially periodic traveling pulse solution. This solution is very similar to the corresponding solution emerging in the $H/L\!\to\!\infty$ limit, extensively discussed in~\cite{Roch2022}.
%%%%%%%%%%%%%%%%%%%%%%%%%%%%%%%%%%%%%%%%%%%%%%%%%%%%%%%%%%%%%%%%%%%%%%%%%%%%%%%%%
%%%%%%%%%%%%%%%%%%%%%%%%%%%%%%%%%%%%%%%%%%%%%%%%%%%%%%%%%%%%%%%%%%%%%%%%%%%%%%%%%

The results presented in this section (and the accompanying results in Sect.~\ref{sec:varying_parameters}) might be rationalized based on the effect of the friction parameter variations and of changing the state evolution law on the elasto-frictional length $L_{\rm c}(H)$. The latter is known to increase with increasing $D$ and with decreasing magnitude of frictional rate-weakening, as quantified by $|df_{\rm ss}(v_0)/d\log\!{(v)}|$~\cite{Aldam2017} (it is also an increasing function of $H$, see Sect.~\ref{sec:SM_Lc}). The parameter variations in Sect.~\ref{sec:varying_parameters}, either reducing $D$ in Fig.~\ref{fig:figS10} or increasing $|df_{\rm ss}(v_0)/d\log\!{(v)}|$ in Figs.~\ref{fig:figS11}-\ref{fig:figS12}, decrease $L_{\rm c}(H)$, which tends to facilitate complexity as observed. Replacing the aging evolution law by the slip evolution law (without changing the frictional parameters), as done in Fig.~\ref{fig:fig4}, leaves $L_{\rm c}(H)$ unchanged. That is, while the two evolution laws differ in their nonlinear dynamics, they share the same linear perturbation response (see, for example,~\cite{Aldam2017}). This independence of $L_{\rm c}(H)$ on the form of the state evolution law (for the two laws considered above) may be related to the observation that complexity emerges for both laws, for the friction parameters used.

Finally, decreasing $|df_{\rm ss}(v_0)/d\log\!{(v)}|$ in Fig.~\ref{fig:fig5} leads to an increase in $L_{\rm c}(H)$. Specifically, for the reference parameters of Table~\ref{tab:SM_parameters} and Fig.~\ref{fig:SM_01_friction_law} (with $H\=0.1$ m, hence corresponding to Fig.~\ref{fig:fig1}) one has $L_{\rm c}(H)\!\simeq\!0.35$ m, see Fig.~\ref{fig:SM_02_nucleation}. This corresponds to $L_{\rm c}(H)/H\!\simeq\!3.5$. On the other hand, the smaller $|df_{\rm ss}(v_0)/d\log\!{(v)}|$ in Fig.~\ref{fig:fig5} corresponds to $L_{\rm c}(H)\!\simeq\!0.7$ m, leading to $L_{\rm c}(H)/H\!\simeq\!7$. One can speculate that this increase in the ratio $L_{\rm c}(H)/H$ might be related to the reverse complexity-to-coarsening transition, i.e., to the disappearance of complexity in Fig.~\ref{fig:fig5}. We also performed simulations with the friction law of Fig.~\ref{fig:fig5}, but with $H\=0.2$ m (i.e., doubling $H$) and did not observe complexity (not shown). In this case, we found $L_{\rm c}(H)\!\simeq\!0.9$ m, leading to $L_{\rm c}(H)/H\!\simeq\!4.5$, which appears to be consistent with the above speculation. If the latter is valid, then complexity in our problem emerges when $H$ becomes sufficiently smaller than $L$ and disappears when it becomes sufficiently smaller than $L_{\rm c}(H)$. Yet, we should note that the results under discussion were all obtained with $L\=4$ m, such that $L/L_{\rm c}(H)$ is not huge and may play a role as well.

\vspace{0.25cm}
%\section*{Discussion}
\hspace{-0.35cm}{\bf \Large Summary and Discussion}

In this work, we studied the effect of a finite height-to-length ratio $H/L$ on the continuum-level dynamics and statistics of unstable, strictly homogeneous rate-and-state dependent frictional systems in 2D. The strictly homogeneous systems feature no quenched disorder and/or geometric irregularities of any form, and their unstable nature --- inherited from frictional rate-weakening --- implies that no spatially homogeneous steady state at the driving velocity $v_0$ exists over long times. It has been recently shown that for $H/L\!\to\!\infty$, such systems initially exhibit complex spatiotemporal dynamics, but as time progresses, they undergo continuous coarsening over increasingly longer lengthscales, until settling into a spatially periodic traveling solution in the form of a steadily propagating pulse~\cite{Roch2022}. From a statistical perspective, this spatially periodic traveling solution implies that all relevant observables are essentially $\delta$-function distributed.

We showed, as implied by earlier reports~\cite{horowitz1989slip,shaw2000existence}, that sufficiently reducing $H/L$ leads to a coarsening-to-complexity transition. In particular, we showed that for $H/L\!\sim\!{\cal O}(0.01\!-\!0.1)$ coarsening is less effective such that the periodic solution is dynamically avoided, and the system settles into a stochastic, statistically stationary state that is characterized by nontrivial statistical distributions. Better understanding the properties of the stochastic steady state and the ways by which the finite geometry gives rise to complexity is of importance. These are relevant both for elucidating physical processes that contribute to complexity in frictional systems and in a broader statistical physics context, for shedding light on the spontaneous self-generation of complexity in physical systems lacking quenched disorder.

Our findings contribute to this effort in four main respects. First, we showed that the slip bursts can be classified into two types (Fig.~\ref{fig:fig1}): small, predominantly non-propagating (aseismic) bursts that are power-law distributed and large, propagating bursts that are log-normally distributed. Both types of slip bursts/events feature a spatial pattern characteristic of rupture (crack-like slip function), yet the small ones do not significantly increase their length during an event. The small slip events are responsible for a non-negligible fraction of the overall slip budget of the system and are shown to play dynamic roles in preparing the interface for nucleating large, propagating slip bursts/events.

The power-law distributions of small, non-propagating events (with an exponent in the range between $-2$ and $-2.5$, slightly dependent on the physical observable, cf.~Fig.~\ref{fig:fig1}) should not be confused with power-law distributions of small earthquakes observed in geophysical contexts (e.g., the Gutenberg-Richter law of earthquake magnitude~\cite{Gutenberg1944}, which corresponds to propagating rupture). Yet, the power-law distributions of small slip events in our system provide clear evidence for the spontaneous emergence of broadly-distributed interfacial heterogeneity and dynamical bulk noise, associated with chaotic wave motion.

Second, we demonstrated how the non-equilibrium, history-dependent dynamics of the frictional interface coupled to the elastodynamics of the finite height bulks give rise to heterogeneity, which in turn affects the nucleation and arrest of slip events (Fig.~\ref{fig:fig2} and Supplemental Movies, see Sect.~\ref{sec:movies}). The emerging heterogeneous slip history of past events is encoded in the state field $\phi(x,t)$ and stress field $\tau(x,t)$. The resulting nontrivial spatial distributions of these interfacial fields --- including aging and other interfacial restrengthening processes --- significantly affect subsequent slip events, which give rise to persistent heterogeneity/disorder. In particular, we demonstrated how the self-generated heterogeneity and spatial gradients affect rupture nucleation in ways that differ from the homogeneous nucleation scenario associated with the elasto-frictional length $L_{\rm c}(H)$. Moreover, the emerging spatial fluctuations appear to be large enough to influence rupture arrest.

Third, we demonstrated the effect of the long strip configuration, $H/L\!\ll\!1$, on frictional rupture dynamics and highlighted the role of the inertial timescale $\sim\!H/c_{\rm s}$ (Fig.~\ref{fig:fig3}a-b). The long strip configuration is relevant to many engineering and geophysical problem, where in the latter context the finite seismogenic zone commonly plays an analogous role to the strip height $H$ (e.g.,~\cite{Weng2017,Weng2019}). $H/c_{\rm s}$ is associated with shear waves that are radiated by frictional rupture and are reflected back from the finite boundary at $y\=\pm H$. We showed that the finite geometric scale $H$ systematically affects the statistics of large events and that reducing the elastic wave-speeds, $c_{\rm s}$ in particular, has a similar effect on the statistics as increasing $H$ (Fig.~\ref{fig:fig3}c-d). It indicates that the wave reflection timescale $\sim\!H/c_{\rm s}$ that carries information regarding the finite geometry is indeed of importance. At the same time, the statistics of the small slip event are weakly dependent on $H$ and $c_{\rm s}$ (Fig.~\ref{fig:fig3}c-d).

Fourth, we showed that slip complexity emerges rather generically within a range of frictional parameters corresponding to sufficiently strong frictional rate-weakening for a fixed $H/L$. Moreover, we demonstrated that within this range of parameters, slip complexity is also largely independent of the state evolution law, whether it is the aging evolution law of Eq.~\eqref{eq:aging_law} or the slip evolution law of Eq.~\eqref{eq:slip_law}. We then demonstrated that by sufficiently reducing the level of frictional rate-weakening, a reverse transition --- upon which complexity disappears --- can be induced. In this case, the spatially periodic traveling pulse solution, observed in the $H/L\!\to\!\infty$ limit~\cite{Roch2022}, is recovered.

It is interesting to note that our findings are somewhat reminiscent of the recent laboratory observations of~\cite{Rubino2022}, where repeated slip events in a frictional system composed of elastic blocks coupled along a contact gouge layer have been probed. While the statistics of slip events are not reported therein and the role of finite boundaries is not explicitly discussed/highlighted, the observations of complex, intermittent slip processes appear to bear some similarities to our observations. In particular, the complex spontaneous sequences of slip events reported in~\cite{Rubino2022} --- involving repeated arrest of rupture propagation, significant rate-and-state dependent frictional strength evolution (stronger than in our system), as well as wave-mediated stress transfer and re-triggering --- echo some of the processes that we identified as underlying slip complexity.

Finally, we note that the transition between the coarsening-mediated, spatially periodic traveling state in the $H/L\!\to\!\infty$ limit (at long times) and the stochastic, statistically stationary state for sufficiently small $H/L$ has not been fully quantified since our cutting-edge, large-scale FEM simulations are computationally prohibited at large $H$ values. While our calculations clearly show that the latter state is realized for $H/L\!\ll\!1$, in agreement with~\cite{horowitz1989slip,shaw2000existence}, we do not know at present the precise transition condition. In particular, we do not know whether the transition occurs at $H\!\sim\!L$ or at $H\!\sim\!L_{\rm c}(H)$. Moreover, we provided evidence indicating that for $H\!\ll\!L_{\rm c}(H)$ a reverse transition in which complexity disappears takes place. Consequently, it seems that for a fixed $L$ (and friction law), there is no complexity for $H/L\!\to\!\infty$, then complexity emerges as $H$ is reduced below some threshold and it disappears once $H$ is sufficiently further reduced. Future work should shed additional light on these issues.

\vspace{0.3cm}
%\section*{Methods}
\hspace{-0.3cm}{\bf Acknowledgements}

E.B.~is supported by the Israel Science Foundation (ISF grant no.~1085/20), the Minerva Foundation (with funding from the Federal German
Ministry for Education and Research), the Ben May Center for Chemical Theory and Computation, and the Harold Perlman Family. T.R.~and E.B.~acknowledge an international cooperation grant (associated with the above-mentioned ISF grant no.~1085/20), which supported a long visit of T.R.~to Weizmann Institute of Science. We thank Mathias Lebihain for developing the data sampling methodology and the identification of slip events.

\vspace{0.2cm}
\hspace{-0.3cm}{\bf Author contributions}

T.R., J.-F.M.~and E.B.~conceived of the project. T.R.~performed all simulations, analyzed the results and generated the figures, with help from E.B.~and J.-F.M., who supervised the research. All authors discussed the results, and their interpretation and significance. T.R.~and E.B.~wrote the manuscript, E.A.B. and J.-F.M. commented on it.\\

\clearpage

\onecolumngrid
%\vspace{1cm}
\begin{center}
              \textbf{\Large Supplemental materials}
\end{center}

%%%%%%%%%%%%%%%%%%%%%%%%%%%%%%%%%%%%%%%%%%%%%%%%%%%%%%%%%%%%%%%%%%%%%%%%%%%%%%%%%
%%%%%%%%%%%%%%%%%%%%%% these lines of code handle the concatenation %%%%%%%%%%%%%
%%%%%%%%%%%%%%%%%%%%%%%%%%%%%%%%%%%%%%%%%%%%%%%%%%%%%%%%%%%%%%%%%%%%%%%%%%%%%%%%%
\setcounter{equation}{0}
\setcounter{figure}{0}
\setcounter{section}{0}
\setcounter{subsection}{0}
\setcounter{table}{0}
\setcounter{page}{1}
\makeatletter
\renewcommand{\theequation}{S\arabic{equation}}
\renewcommand{\thefigure}{S\arabic{figure}}
\renewcommand{\thesection}{S-\Roman{section}}
\renewcommand*{\thepage}{S\arabic{page}}
%\renewcommand{\bibnumfmt}[1]{[S#1]}
%\renewcommand{\citenumfont}[1]{S#1}
%%%%%%%%%%%%%%%%%%%%%%%%%%%%%%%%%%%%%%%%%%%%%%%%%%%%%%%%%%%%%%%%%%%%%%%%%%%%%%%%%
%%%%%%%%%%%%%%%%%%%%%% these lines of code handle the concatenation %%%%%%%%%%%%%
%%%%%%%%%%%%%%%%%%%%%%%%%%%%%%%%%%%%%%%%%%%%%%%%%%%%%%%%%%%%%%%%%%%%%%%%%%%%%%%%%
\twocolumngrid

In this Supplemental materials file, we provide additional technical details regarding the friction law, the numerical methodology and the slip events identification procedure. In addition, we present some additional results that support statements made in the manuscript.

\section{The friction law}
\label{sec:SM_friction_law}

\begin{figure}[ht!]
    \centering
    {\includegraphics[]{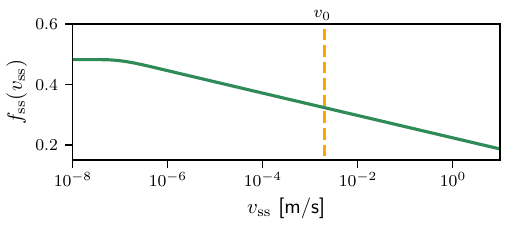}}
    \caption{The steady-state frictional strength $f_{\rm ss}(v_{\rm ss})\!\equiv\!\tau[v_{\rm ss},\phi_{\rm ss}(v_{\rm ss})]/\sigma_0$ vs.~the steady-state slip velocity $v_{\rm ss}$, presented on a semi-logarithmic scale. The vertical dashed line corresponds to the driving velocity used, $v_0$.}
    \label{fig:SM_01_friction_law}
\end{figure}

The interfacial constitutive law is formulated within a rate-and-state friction framework \cite{Baumberger2006,Ruina1983,Marone1998a,Nakatani2001}, see manuscript for an extended list of relevant references. In particular, the frictional strength we employed, which is based on extensive laboratory data~\cite{Baumberger2006,Ruina1983,Marone1998a,Nakatani2001,Barsinai2014}, takes the form
\begin{equation}
\label{eq:6_fem}
\tau(v,\phi)/\sigma_0\!=\! f_0 \!\left[1 + \!b \log \left(\!\frac{\phi}{\phi_*}\!\right)\!\right] \!+\! a \log\left(\!1\!+\!\frac{|v|}{v_*}\!\right) \ ,
\end{equation}
with $v$ being the slip velocity and $\phi$ the state variable, see manuscript for discussion. $\sigma_0$ is the normal compressive stress acting on the interface and $a,b,D,v_*,\phi_*$ are rate-and-state parameters, whose values are specified in Table~\ref{tab:SM_parameters}. The state variable obeys the aging evolution law~\cite{Dieterich1979}
\begin{equation}
\label{eq:6_evolution_law}
\partial_t\phi = 1  - \frac{|v|\phi}{D}\sqrt{1 + (\Vstar/v)^2} \ ,
\end{equation}
where the square-root factor on the right-hand-side ensures that $\phi$ saturates, rather than diverges, for very small steady-state velocities $v_{\rm ss}$. This factor, i.e., $\sqrt{1 + (\Vstar/v)^2}$, is not included in Eq.~\eqref{eq:aging_law} as it does not affect our results beyond the mention regularization of steady-state friction. The resulting steady-state curve (corresponding to $\partial_t\phi\=0$, leading to $\phi_{\rm ss}(v_{\rm ss})$, which reads $\phi_{\rm ss}\!\simeq\!D/v_{\rm ss}$ for $v_{\rm ss}\!\gg\!\Vstar$) is rate (velocity) independent at very small slip velocities and velocity-weakening (rate-weakening) otherwise. This is demonstrated in Fig.~\ref{fig:SM_01_friction_law}, where $f_{\rm ss}(v_{\rm ss})\!\equiv\!\tau[v_{\rm ss},\phi_{\rm ss}(v_{\rm ss})]/\sigma_0$ is plotted. The values of the friction law parameters are presented in Table~\ref{tab:SM_parameters}. The latter also provides the mass density $\rho$, the shear modulus $\mu$ and Poisson's ratio $\nu$.

\begin{table}[ht!]
  \begin{center}
    \begin{tabular}{p{3cm} p{3cm} r}
      \hline
      \hline
      Parameter & Value & Unit \\
      \hline
      $\nu$ & 0.33 & ... \\
      $\mu$ & $9\times10$\textsuperscript{9} & Pa \\
      $\rho$ & 1200 & kg/m\textsuperscript{3} \\
      $f_0$ & 0.28 & ... \\
      $a$ & 0.005 & ... \\
      $b$ & 0.075 & ... \\
      $D$ & 5 $\times$ 10\textsuperscript{-7} & m \\
      $v_{*}$ & 1 $\times$ 10\textsuperscript{-7} & m/s \\
      $\phi_{*}$ & 3.3 $\times$ 10\textsuperscript{-4} & s \\
      \hline
      \hline
    \end{tabular}
  \end{center}
  \caption{The values of bulk and rate-and-state friction parameters used in this work, unless otherwise stated. Note that $\rho$ has been varied between 300 and 4800 kg/m\textsuperscript{3} in Fig.~\ref{fig:fig3}d, in order to study the effect of varying the elastic wave-speeds. Specific variations of friction parameters are discussed in the main text and in Sect.~\ref{sec:varying_parameters}.
    \label{tab:SM_parameters}}
\end{table}

\section{The elasto-frictional nucleation length $L_{\rm c}(H)$}
\label{sec:SM_Lc}

\begin{figure}[ht!]
    \centering
    \includegraphics[]{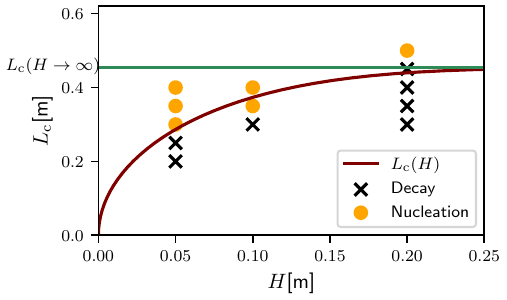}
    \caption[Theoretical and observed nucleation lengths in finite systems.]{The elasto-frictional nucleation length $L_{\rm c}(H)$. The theoretical nucleation length, obtained in a linear stability analysis following~\cite{Aldam2017}, is shown in brown. The horizontal green line indicates the theoretical nucleation length in the limit of $H\!\to\!\infty$. The results of a numerical stability analysis (see text for details) agree with the theoretical prediction (instability is indicated by the orange circles and stability by the black crosses), and demonstrate the two-dimensional nature of our system (also see text for details).}
    \label{fig:SM_02_nucleation}
\end{figure}

The rate-weakening ($df_{\rm ss}/dv_{\rm ss}\!<\!0$) nature of the steady-state friction law implies the existence of a linear elasto-frictional instability~\cite{Rice1983,Rice1996,Aldam2017}. The latter is characterized by a critical nucleation length $L_{\rm c}(H)$. We aim to verify that the systems under consideration fully resolve two-dimensional dynamics (i.e., are not in the quasi-one-dimensional limit~\cite{Aldam2017}). To this end, we computed the theoretical nucleation length through a linear stability analysis for two homogeneous symmetric elastic bodies of height $H$ in contact along a planar interface~\cite{Aldam2017}. The resulting $L_{\rm c}(H)$ is shown (solid brown line) in Fig.~\ref{fig:SM_02_nucleation}, while the horizontal green line indicates the $H\!\to\!\infty$ limit. For the case discussed in the manuscript with $H\=0.1$ m, the critical nucleation length corresponds to $\approx 80\%$ of $L_{\rm c}(H\to \infty)$, supporting the hypothesis of two-dimensional dynamics. $L_{\rm c}$ varies from $\approx 65\%$ of $L_{\rm c}(H\!\to\!\infty)$ (with $H\=0.05$ m) to almost $\approx 100\%$ of $L_{\rm c}(H\!\to\!\infty)$ (with $H\=0.2$ m).

To test the validity of the theoretical estimate for $L_{\rm c}(H)$ in our calculations, we numerically studied the stability of our frictional system. In particular, we conducted finite element simulations with an initial sinusoidal perturbation in the state field of the form
\begin{equation}
    \delta \phi(x,t=0) = \phi_{\rm ss}\,\epsilon\,\sin\left( 2\pi x / L + \pi / 2 \right) \ ,
\end{equation}
with $\epsilon\=1\times\!10$\textsuperscript{-4}.

We report in Fig.~\ref{fig:SM_02_nucleation} the outcome of this numerical perturbation analysis. The perturbation either decays (black crosses) or grows (orange circles), where the latter leads to rupture nucleation. The numerical results are in good quantitative agreement with the theoretical prediction, further supporting the two-dimensional nature of the system studied. This implies that singular rupture fronts can fully develop, contrarily to the quasi-one-dimensional limit~\cite{Aldam2017}.

As explained in the manuscript, since the driving velocity $v_0$ resides on the rate-weakening part of the steady-state friction curve (cf.~Fig.~\ref{fig:SM_01_friction_law}) and in view of the relation $L_{\rm c}\!\ll\!L$, a linear elasto-frictional instability would be spontaneously triggered (e.g., by numerical noise alone). Yet, as we are interested in the long-time, stationary dynamics of the system, we initially perturb the interface by adding spatial Gaussian noise to the state variable.

\section{The numerical method}
\label{sec:SM_fem}

The system sketched in Fig.~1a in the manuscript is studied in the limit of $w\!\ll\!H,L$, i.e., a two-dimensional configuration. Two identical bodies, of length $L$ and height $H$ each, are in contact along a planar interface. The top and bottom boundaries are driven by a prescribed velocity $v_0/2$ in the $x$ direction and by a compressive stress of magnitude $\sigma_0\=1\times10\textsuperscript{6}$ Pa. The bodies are initially moving uniformly in opposite directions at $v_0/2$. Periodic boundary conditions are enforced at the lateral edges $x\=0$ and $x\=L$ (these periodic boundary conditions in the sliding direction make the system equivalent to the rotary system illustrated in Fig.~1a in the manuscript). The interface is initially at steady-state with $v_{\rm ss}\=v_0$ and $\phi_{\rm ss}=D/v_0$.

Our calculations are performed using the explicit dynamic finite element framework, based on an in-house open-source finite element library called Akantu~\cite{Richart2015}. The domain is discretized into a regular mesh composed of bilinear quadrilateral elements (Q4). The sliding interface between the two elastic bodies is modeled using a node-to-node contact algorithm, see~\cite{Rezakhani2020} for details. Time integration is performed using the central difference method and the time step is taken small enough to eliminate the numerical instabilities associated with the explicit finite element modeling of rate-and-state friction, as explained in~\cite{Rezakhani2020}. In our simulations, we set the time step to $\Delta{t}_{\rm F}\=\alpha_{\rm F} \Delta{t}_{\rm CFL}$, where $\Delta{t}_{\rm CFL}$ is determined by the Courant-Friedrichs-Lewy condition and $\alpha_{\rm F}$ is typically taken to be ${\cal O}(0.01)$.

\subsection{Mesh discretization}

The typical mesh size is chosen such that both the elasto-frictional nucleation length $L_{\rm c}(H)$ (discussed above, see Sect.~\ref{sec:SM_Lc}) and the process zone size $\Delta_{\rm pz}$ at the edge of propagating frictional rupture are properly resolved. The former ensures that the system is not intrinsically discrete~\cite{Rice1993,Ben-zion1993,Ben-zion1995,Rice1996}, while the latter is required to faithfully account for slip dynamics. In a typical simulation, we set $\Delta{x}\=2.5\times\!10$\textsuperscript{-3} m such that the nucleation length is discretized with $L_{\rm c}/\Delta{x}\= \mathcal{O}(100)$ elements. The process zone size can be estimated as  $\Delta_{\rm pz}\=\mathcal{O}(\mu D /b\,\sigma_0\,[1-\nu])\=9.5\times\!10$\textsuperscript{-2} m, see~\cite{Lapusta2009_3D}, implying that is it resolved by $\sim\!40$ elements.

\section{The identification of slip events}
\label{sec:SM_identification}

Here, we summarize the procedure we employed to identify distinct slip events and effectively the operational definition of the latter. The simulation taken as an example here is not the one discussed in the manuscript (shown in Fig.~1 therein), but rather a shorter duration one with $H\=0.2$ m. This choice is purely presentational, motivated by visual clarity considerations (the very same procedure for events identification was applied to all of our simulations).

\begin{figure*}[ht!]
  \centering
  {\includegraphics[width=1\textwidth]{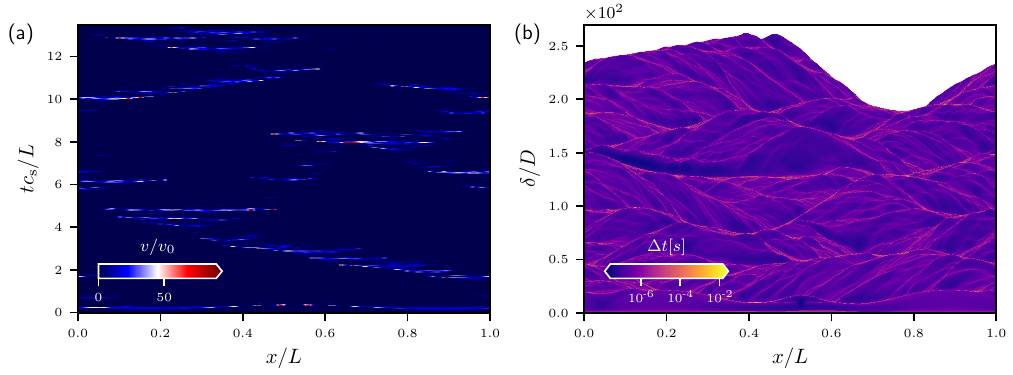}}
  \caption{(a) Space-time plot of the velocity $v(x,t)$ during a typical simulation, normalized by the driving velocity $v_0$. The part of the plot in dark blue corresponds to the interface sticking. (b) Space-slip plot of the time required to slip over a distance $\delta_{\rm th}$. The lower values in dark purple indicate a rapid slip, while the largest ones in bright yellow indicate a long waiting period, i.e., sticking conditions. The simulation shown here corresponds to $L\!=\!4$ m, $H\!=\!0.2$ m, $v_0\!=\!2\times\!10$\textsuperscript{-3} m/s.}
  \label{fig:SM_03_ct_slipmap}
\end{figure*}

The slip activity on the fault is expected to be strongly heterogeneous, both in time and space (fast, spatially localized slip events, followed by long waiting periods), see an example in Fig.~\ref{fig:SM_03_ct_slipmap}a. Most of the time, the interface is sticking (dark blue in Fig.~\ref{fig:SM_03_ct_slipmap}a). The focus is on the slip events, and thus we use a slip interval instead of a time interval for sampling purposes. This method is used in the context of depinning physics (see, for example,~\cite{lebihain2019}). Using slip sampling allows us to obtain detailed information during slip events and to reduce the amount of data corresponding to the waiting time between slip events, where the slip velocity essentially vanishes.

To proceed, we define a slip threshold $\delta_{\rm th}$, which will eventually allow us to define discrete slip events, as explained next. We use $\delta_{\rm th}\=10^{-8}\,\hbox{m}\=0.02\,D$ (cf.~Table~\ref{tab:SM_parameters} for the latter) in this work. We record the time it took a point along the interface to slip from $\delta\= 0$ to $\delta\= \delta_{\rm th}$, then from $\delta\=\delta_{\rm th}$ to $\delta\=2 \delta_{\rm th}$ and so on. The outcome of this procedure is called a space-slip map of the duration $\Delta t(x,\delta)$, which represents the time it took an interfacial point $x$ to accumulate slip from $\delta\!-\!\delta_{\rm th}$ to $\delta$. An example is shown in Fig.~\ref{fig:SM_03_ct_slipmap}b, with the corresponding space-time plot of the velocity presented in Fig.~\ref{fig:SM_03_ct_slipmap}a. The slip is normalized by $D$, the characteristic slip distance of the rate-and-state friction framework. In this space-slip map, the purple patches correspond to short duration $\Delta t$ required to slip a given $\delta_{\rm th}$, i.e., a large sliding velocity, while the bright orange/yellow areas correspond to stick conditions. This map contains significantly more information than the one shown in Fig.~\ref{fig:SM_03_ct_slipmap}a.

\begin{figure*}[ht!]
  \centering
  {\includegraphics[width=1\textwidth]{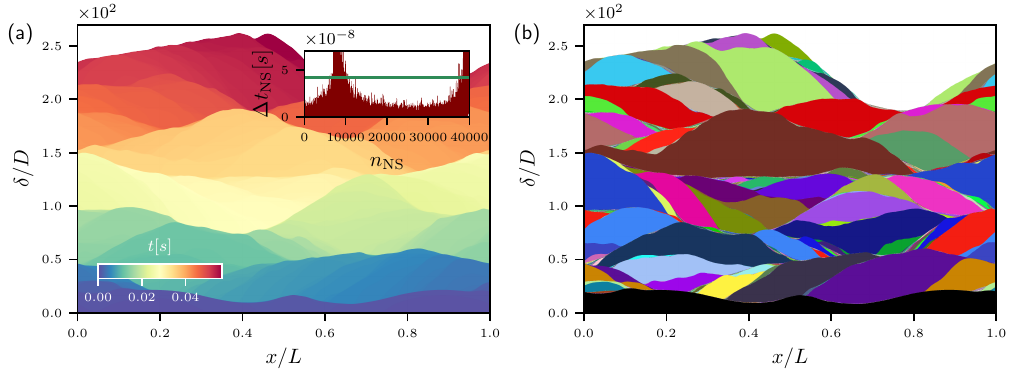}}
  \caption{(a) Space-slip map of the time $t$ at which a given cumulative slip value $\delta/D$ has been reached at a given position on the interface $x/L$. An area sharing the same color corresponds to a slip that occurred at the same time, while discontinuity in the colors indicates a waiting period. (inset) An extract of the waiting time $\Delta t_{\rm NS}$ between NSs. The green horizontal line corresponds to the time threshold $\alpha\,\delta_{\rm th}/(v_0 N)$ with $\alpha\!=\!15$ (see text for the other parameters). (b) The corresponding space-slip map of the distinct slip events, see text for more details on the procedure. The simulation shown here corresponds to $L\!=\!4$ m, $H\!=\!0.2$ m, $v_0\!=\!2\times\!10$\textsuperscript{-3} m/s.}
  \label{fig:SM_04_time_avalanche}
\end{figure*}

The time $t(x,\delta)$ at which a given cumulative slip threshold value has been reached, is directly obtained as the sum of the duration for all lower slip thresholds. Figure~\ref{fig:SM_04_time_avalanche}a shows the space-slip time map obtained from Fig.~\ref{fig:SM_03_ct_slipmap}b. The content of the space-slip time map, which in view of the discretized nature of $x$ and $\delta$ is in fact a matrix, is then lumped together into a vector ${\bm t}_{\rm NS}$. As we are interested in slip events, which might simultaneously occur at different spatial locations, we sort ${\bm t}_{\rm NS}$ by time of occurrence. The result contains the information on when a single numerical node has reached a new slip threshold, which we refer to as a nodal slip (NS). We then compute the time interval between two successive nodal slips anywhere on the interface $\Delta t_{\rm NS}$. If the time interval is sufficiently large (i.e., larger than a given threshold), it means that the interface is entirely sticking and thus indicates the separation between two independent slip events. The waiting time threshold is chosen taking into account the number of interfacial nodes, the driving velocity, and the slip threshold, as explained next.

In the framework of the adopted procedure, the time interval between NSs decreases with the number of nodes $N$ we use to discretize the interface. Consequently, in selecting the above-mentioned waiting time threshold, we take into account both the characteristic time $\delta_{\rm th}/v_0$ and $N$. In practice, we set our waiting time threshold to a multiple of $\delta_{\rm th}/(v_0N)$, i.e., $\alpha\,\delta_{\rm th}/(v_0 N)$, with $\alpha$ being a dimensionless constant of ${\cal O}(10)$. We used $\alpha\=15$ throughout this work, which is chosen such that it has a weak effect on the emerging statistical distributions of the slip events, yet allowing the procedure to reasonably distinguish between slip events, see Sect.~\ref{sec:SM_threshold_effect}. The typical value for the number of nodes is $N\=1600$.

An extract of the time interval vector $\Delta t_{\rm NS}$ taken at a given time is illustrated in the inset of Fig.~\ref{fig:SM_04_time_avalanche}a, where $n_{\rm NS}$ is the number of NSs. Note that in the entire simulation shown in Fig.~\ref{fig:SM_04_time_avalanche}a, there are $\mathcal{O}(10^7)$ NSs. The waiting time threshold is indicated in the inset of Fig.~\ref{fig:SM_04_time_avalanche}a by the horizontal green line. All the NS that falls in between two values exceeding the threshold (for example, from $n_{\rm NS}\!\simeq\!10000$ to $n_{\rm NS}\!\simeq\!38000$ in the inset of Fig.~\ref{fig:SM_04_time_avalanche}a) are grouped inside a single \emph{slip event}. This information is then digitized back on the space-slip map, and we obtain a visual description of the distinct slip events as shown in Fig.~\ref{fig:SM_04_time_avalanche}b, in which each color corresponds to a slip event. There are $\mathcal{O}(10^4)$ slip events after applying this procedure in Fig.~\ref{fig:SM_04_time_avalanche}b. Note that slip events that are too small, both spatially $\mathcal{O}(\Delta{x})$ and in terms of slip $\mathcal{O}(\delta_{\rm th})$, are removed from the analysis.

\begin{figure}[ht!]
  \centering
  {\includegraphics[width=0.5\textwidth]{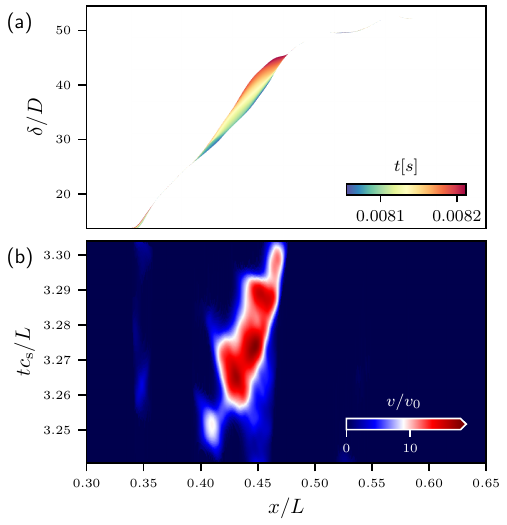}}
  \caption{(a) Space-slip map of the time $t$ at which a given cumulative slip value $\delta / D$ has been reached at a given position on the interface $x/L$ for a single event. Only the colored area is part of the event. (b) The space-time velocity map of this event. It was reconstructed from the information in the space-slip map.}
  \label{fig:SM_05_example}
\end{figure}

An example of a single slip event is shown in Fig.~\ref{fig:SM_05_example}a in the space-slip map of the time, with the corresponding space-time map of the velocity shown in Fig.~\ref{fig:SM_05_example}b. The latter has been reconstructed from the information stored in the space-slip map of the duration and is not directly sampled in time. The red patch in Fig.~\ref{fig:SM_05_example}b, spanning from $x/L\=0.4$ to $0.47$ is the main rupture area with a sliding velocity going up to $\mathcal{O}(10\,v_0)$. This rupture is growing and propagating spatially. Around $x/L\=0.35$, there is some non-sticking velocity (light blue), but these are not propagating events and slip is most likely triggered by wave reflections in the bulk, generated by past slip events and interactions with the finite boundaries.

\begin{figure}[ht!]
  \centering
  {\includegraphics[width=0.5\textwidth]{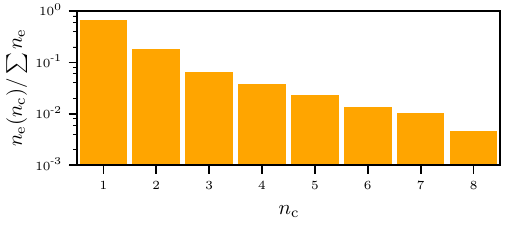}}
  \caption[Distribution of the number of clusters per slip event.]{The number of slip events $n_{\rm e}(n_{\rm c})$ composed of $n_{\rm c}$ clusters, normalized by the total number of events. Note the semi-logarithmic scale.}
  \label{fig:SM_06_clusters}
\end{figure}

We also investigated the structure of the slip events, i.e., assess if they are spatially connected or composed of spatially disconnected slipping patches, which we call here \emph{clusters}. Once again, the smallest clusters, corresponding to slip of $\mathcal{O}(\delta_{\rm th})$ or to spatial extent of $\mathcal{O}(\Delta{x})$, are filtered out. The distribution of the number of clusters $n_{\rm c}$ per slip event is shown in Fig.~\ref{fig:SM_06_clusters}. It is observed that most events are composed of a single cluster (note the semi-logarithmic scale).

Once the slip events are identified, we can compute their characteristics: the spatial extent of the rupture $L_{\rm s}$, duration $T$, average slip $\overline{u}$, seismic moment $M_0 = \mu L_{\rm s} \overline{u}$ with $\mu$ the shear modulus of the bulk surrounding the interface. Note that the first and last events in each simulation are excluded from this analysis. The first one is always spanning the entire interface and is triggered by the initial noise, while the last one is potentially not complete when the simulation ends.

\section{Additional supporting results}
\label{sec:additional_results}

\subsection{Statistical stationarity}
\label{sec:SM_stationarity}

\begin{figure*}
  \centering
    \includegraphics[width=\textwidth]{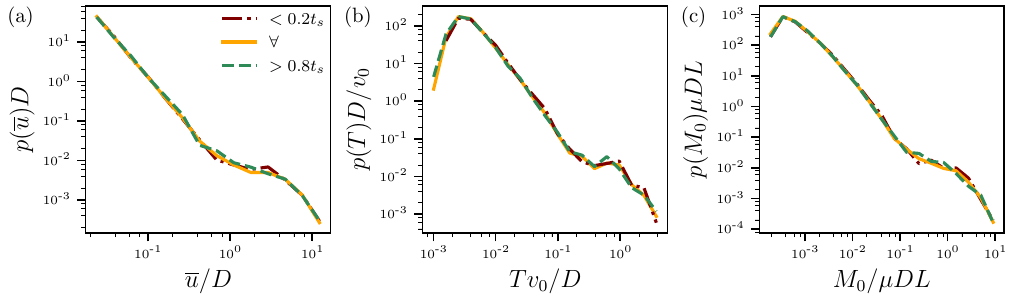}
  \caption{The probability density function $p(x)$ of: (a) the average slip $\overline{u}$ (normalized by $D$), (b) the duration $T$ (normalized by $D/v_0$) and (c) the seismic moment $M_{0}$ (normalized by $\mu D L$) for three sets of events: the entire data set (orange), the first fifth of the simulation (dotted dashed brown) and the last fifth (dashed green).}
  \label{fig:SM_08_temporal}
\end{figure*}

In the manuscript, we state that the system reaches a statistical steady-state, where most of our analysis is performed. Here, our goal is to demonstrate this statistical stationarity.
A direct test is obtained by evaluating the distributions of various physical quantities over time. In Fig.~\ref{fig:SM_08_temporal}, we show the probability density functions (PDFs) for the average slip $\overline{u}$ per event, the event duration $T$ and the seismic moment $M_0$ per event for the entire simulation (of total duration $t_{\rm s}$, in orange), and two subsets of events. The latter correspond,  respectively, to the events occurring during the first fifth of the simulation (dotted dashed brown, $t\!<\!0.2 t_{\rm s}$) and to the last fifth of the simulation (dashed green, $t\!>\!0.8 t_{\rm s}$). It is observed that all 3 distributions, for all quantities, are essentially the same, demonstrating statistical stationarity.
\begin{figure*}
  \centering
    \includegraphics[width=\textwidth]{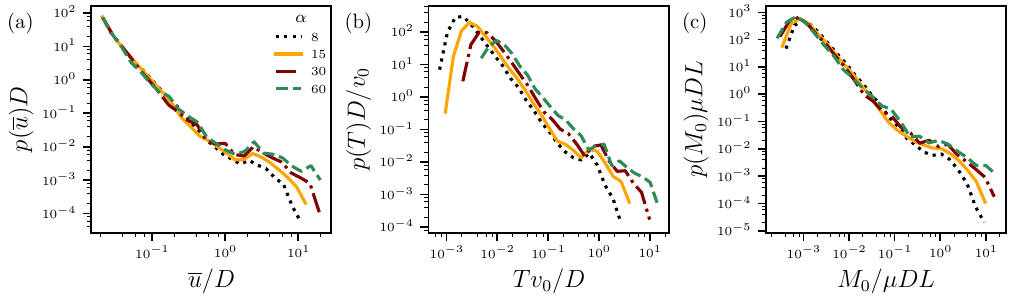}
  \caption[Influence of the arbitrary thresholds on the probability density functions.]{The probability density function $p(x)$ of: (a) the average slip $\overline{u}$ (normalized by $D$), (b) the duration $T$ (normalized by $D/v_0$) and (c) the seismic moment $M_{0}$ (normalized by $\mu D^2$) for four values of the time threshold $\alpha\,\delta_{\rm th}/(v_0 N)$, which is varied by varying $\alpha$ (where other parameters are kept fixed). See legend on panel (a) for the values of $\alpha$ used.}
  \label{fig:SM_09_threshold}
\end{figure*}

\subsection{The effect of the waiting time threshold on the statistical distributions}
\label{sec:SM_threshold_effect}

Here, our goal is to demonstrate that the main results discussed on the manuscript are independent of the chosen time threshold $\alpha\,\delta_{\rm th}/(v_0 N)$ for defining discrete slip events. In Fig.~\ref{fig:SM_09_threshold}, the probability density functions of the average slip, duration, and seismic moment, determined with $\alpha = 8,15,30,60$ are shown.
The main qualitative characteristics of the PDFs discussed in the manuscript are independent of the threshold. First, the distribution of small events follows a  power law scaling. Second, a change in scaling law occurs around the same scale, which is related to the reflection timescale. Finally, the large propagating events are roughly log-normal distributed.

Note that the behaviors near the cut-offs are, however, different (as expected): modifying the threshold directly affects the minimal event size. In addition, increasing $\alpha$ results in grouping events together and thus increases the maximum observed event size. For values significantly larger than shown in Fig.~\ref{fig:SM_09_threshold}, numerous slip events are grouped together up to a point where the entire interface history is considered as a single slip event.

\subsection{Wave reflection timescale and the average slip distribution}
\label{sec:SM_reflection}

\begin{figure*}[ht!]
  \centering
    \includegraphics[width=\textwidth]{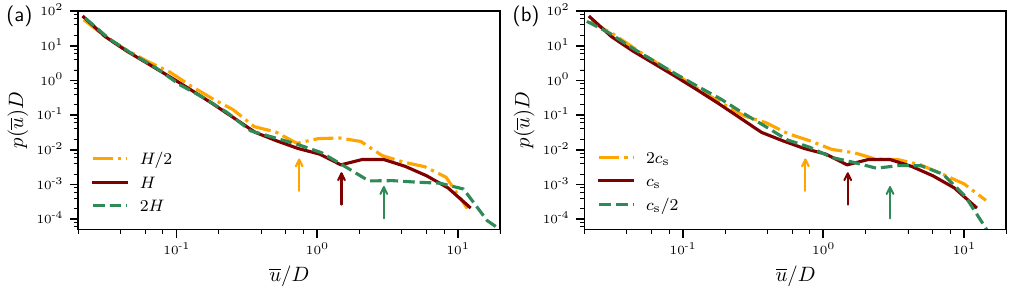}
  \caption{(a) $p(\overline{u})$ for 3 values of $H$ as indicated in the legend, where the reference case is $H\!=\!0.1$ m. The upward-pointing arrows mark the values of
$\eta\,v_0\,2H/c_{\rm s}$ (same color code as in the legend). (b) The same as panel (a), but for 3 values of $c_{\rm s}$ as indicated
in the legend, where the upward pointing arrows mark the values of $\eta\, v_0 2H/c_{\rm s}$. In both panels, we set $\eta\!=\!3$.}
  \label{fig:SM_reflexion}
\end{figure*}

In the manuscript, we discussed the effect of the reflection timescale $2H/c_{\rm s}$ on the distribution of the slip event duration. Here, we show that this timescale manifests itself in the distribution of other quantities, such as the average slip $\overline{u}$. The latter are shown in Fig.~\ref{fig:SM_reflexion}, with the effect of increasing/reducing $H$ in panel (a) and the increasing/reducing $c_{\rm s}$ in panel (b). The peak in the distribution of large events is shifted to the right when increasing the timescale $2H/c_{\rm s}$ in both cases. Note that the arrow indicates a slip scale $\eta\,v_0 2H/c_{\rm s}$, corresponding to a characteristic slip accumulated during the reflection timescale (in Fig.~\ref{fig:SM_reflexion}, we set $\eta\!=\!3$).

\subsection{Robustness against variations in the friction law parameters}
\label{sec:varying_parameters}

\begin{figure*}[ht!]
  \centering
    \includegraphics[width=\textwidth]{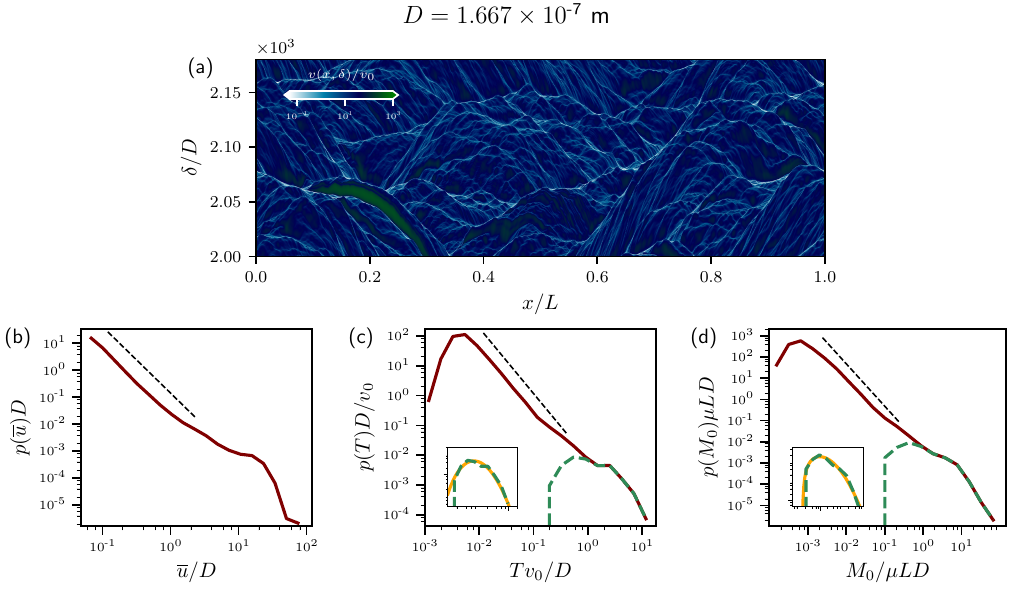}
  \caption{Results of a simulation identical to the one corresponding to Fig.~\ref{fig:fig1}, except for using a slip distance $D$ that is 3 times smaller, set to $D\!=\!1.667\times10$\textsuperscript{-7} m. Decreasing $D$ results in a smaller critical nucleation length $L_{\rm c}$ and consequently leads to a smaller minimal size for propagating events. (a) A space-slip plot of the slip velocity $v(x,\delta)/v_0$ in the $x/L\,$--$\,\delta/D$ plane, qualitatively demonstrating the complexity in the system, similarly to Fig.~\ref{fig:fig1}b. (b) The stationary probability distribution function $p(\bar{u})$. (c) The stationary probability distribution function $p(T)$. The corresponding distribution of slip events featuring $\bar{u}\!>\!D$ is superimposed (dashed green line), similarly to Fig.~\ref{fig:fig1}c. The inset presents a log-normal fit (solid yellow line) to the latter. (d) The stationary probability distribution function $p(M_0)$. The superimposed dashed lines in panels (b), (c) and (d) correspond to power-laws with the same exponents as in Fig.~\ref{fig:fig1}c-e to highlight the quantitative similarity between the two sets of results.}
  \label{fig:figS10}
\end{figure*}

\begin{figure*}[ht!]
  \centering
    \includegraphics[width=\textwidth]{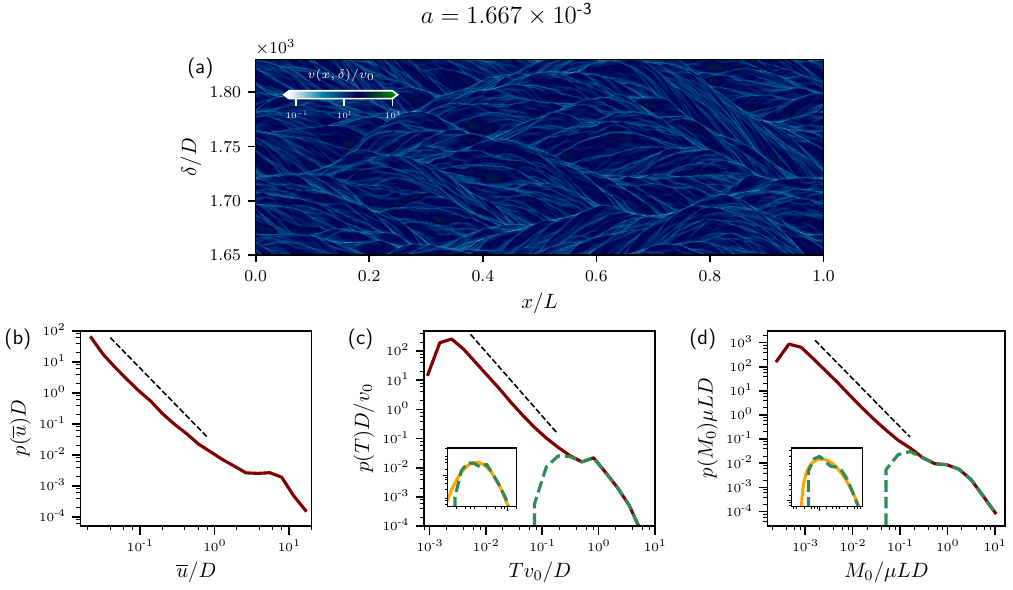}
  \caption{Results of a simulation identical to the one corresponding to Fig.~\ref{fig:fig1}, except for the value of the friction parameter $a$ (quantifying the magnitude of the logarithmic rate dependence in Eq.~\eqref{eq:6_fem}) that is 3 times smaller, set to $a\!=\!1.667\times10$\textsuperscript{-3}. Decreasing $a$ results in stronger rate-weakening behavior at intermediate velocities and in a smaller critical nucleation length $L_{\rm c}$. Note that increasing $a$ has the opposite effect, hence facilitating rupture that propagates over larger distances. (a) A space-slip plot of the slip velocity $v(x,\delta)/v_0$ in the $x/L\,$--$\,\delta/D$ plane, qualitatively demonstrating the complexity in the system, similarly to Fig.~\ref{fig:fig1}b. (b) The stationary probability distribution function $p(\bar{u})$. (c) The stationary probability distribution function $p(T)$. The corresponding distribution of slip events featuring $\bar{u}\!>\!D$ is superimposed (dashed green line), similarly to Fig.~\ref{fig:fig1}c. The inset presents a log-normal fit (solid yellow line) to the latter. (d) The stationary probability distribution function $p(M_0)$. The superimposed dashed lines in panels (b), (c) and (d) correspond to power-laws with the same exponents as in Fig.~\ref{fig:fig1}c-e to highlight the quantitative similarity between the two sets of results.}
  \label{fig:figS11}
\end{figure*}

\begin{figure*}[ht!]
  \centering
    \includegraphics[width=\textwidth]{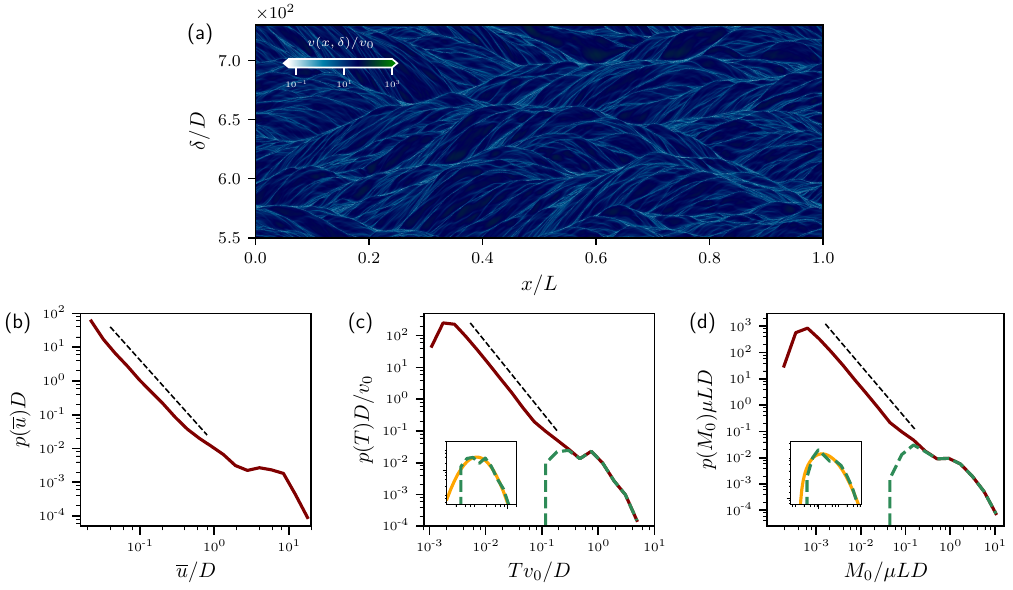}
  \caption{Results of a simulation identical to the one corresponding to Fig.~\ref{fig:fig1}, except that we use $f_0\!=\!0.313$, $a\!=\!0.00166$ and $b\!=\!0.067$, see Eq.~\eqref{eq:6_fem}. The latter are chosen such that their coordinated effect is to maintain $f_{\rm ss}(v_0)$ at its reference value (recall that $v_0$ is the driving velocity), but to increase the magnitude of rate-weakening (i.e., make the slope of the steady-state friction curve more negative). Stronger rate-weakening behavior implies a smaller critical nucleation length $L_{\rm c}$. Note that if one uses a smaller slope for the steady-state friction (i.e., weaker rate-weakening), then it becomes significantly more difficult to observe complexity as the system starts hosting stable rupture in the form of self-healing pulses that propagate over multiple domain sizes.
  (a) A space-slip plot of the slip velocity $v(x,\delta)/v_0$ in the $x/L\,$--$\,\delta/D$ plane, qualitatively demonstrating the complexity in the system, similarly to Fig.~\ref{fig:fig1}b. (b) The stationary probability distribution function $p(\bar{u})$. (c) The stationary probability distribution function $p(T)$. The corresponding distribution of slip events featuring $\bar{u}\!>\!D$ is superimposed (dashed green line), similarly to Fig.~\ref{fig:fig1}c. The inset presents a log-normal fit (solid yellow line) to the latter. (d) The stationary probability distribution function $p(M_0)$. The superimposed dashed lines in panels (b), (c) and (d) correspond to power-laws with the same exponents as in Fig.~\ref{fig:fig1}c-e to highlight the quantitative similarity between the two sets of results.}
  \label{fig:figS12}
\end{figure*}

Following the discussion in the main text, we consider here the effect of varying the rate-and-state friction law parameters, in the framework of its formulation in Eqs.~\eqref{eq:6_fem}-\eqref{eq:6_evolution_law} (see also Eq.~\eqref{eq:aging_law}). In Fig.~\ref{fig:figS10}, we present results of a simulation identical to the one corresponding to Fig.~\ref{fig:fig1}, including all of the parameters listed in Table~\ref{tab:SM_parameters}, except for using a slip distance $D$ that is 3 times smaller (see figure caption). The presentational format of Fig.~\ref{fig:figS10}a-d is the same as that of Fig.~\ref{fig:fig1}b-e. It is observed that the resulting statistical distributions in Fig.~\ref{fig:figS10}b-d are very similar to their counterparts in Fig.~\ref{fig:fig1}c-e. In Fig.~\ref{fig:figS11}, we present results as in Fig.~\ref{fig:fig1}, except that the value of the friction parameter $a$ (quantifying the magnitude of the logarithmic rate dependence in Eq.~\eqref{eq:6_fem}) is 3 times smaller. In Fig.~\ref{fig:figS12}, we present results as in Fig.~\ref{fig:fig1}, except for changing the values of 3 friction parameters, as indicated in the figure caption. Both figures further demonstrate that the emergence of complexity is not affected by these variations in the parameters of the friction law. As discussed and explained in the main text, these variations correspond to decreasing the elasto-frictional length $L_{\rm c}(H)$, either by reducing $D$ (Fig.~\ref{fig:figS10}) or by increasing the magnitude of frictional rate-weakening (Figs.~\ref{fig:figS11}-\ref{fig:figS12}).

\clearpage
\subsection{Supplemental movies}
\label{sec:movies}

Three supplemental movies are associated with this manuscript:
\begin{itemize}
    \item ``Supplemental Movie1.avi'', focusing on ``event 1'' in Fig.~2 in the manuscript.
    \item ``Supplemental Movie2.avi'', focusing on ``event 4'' in Fig.~2 in the manuscript.
    \item ``Supplemental Movie3.avi'', focusing on ``event 5'' in Fig.~2 in the manuscript.
\end{itemize}
Each movie features 4 vertical panels, sharing the same $x$ axis, $x/L$, and progress in time with $t c_{\rm s}/L$, shown on top. The size of the elasto-frictional length $L_{\rm c}$ (also in units of $L$) is presented in panel (a) in each movie. Each movie presents the time evolution of 4 interfacial fields: (a) the slip velocity (normalized by $v_0$), (b) the accumulated slip since the event's onset (normalized by $D$), (c) the state field (in units of seconds) and (d) the interfacial shear stress (normalized by the compressive stress $\sigma_0$). The movies are discussed in the manuscript in relation to Fig.~2 and Fig.~3, and are available \href{https://youtube.com/playlist?list=PLT7c4IN61XQPfKNW-MN_A4oedmtblFf3q&si=kgWBKEeTc6VMjvRO}{online}.

%% Loading bibliography style file
%\bibliographystyle{elsarticle-harv}
% Loading bibliography database
%\bibliography{complexity}

\begin{thebibliography}{71}
\expandafter\ifx\csname natexlab\endcsname\relax\def\natexlab#1{#1}\fi
\providecommand{\url}[1]{\texttt{#1}}
\providecommand{\href}[2]{#2}
\providecommand{\path}[1]{#1}
\providecommand{\DOIprefix}{doi:}
\providecommand{\ArXivprefix}{arXiv:}
\providecommand{\URLprefix}{URL: }
\providecommand{\Pubmedprefix}{pmid:}
\providecommand{\doi}[1]{\href{http://dx.doi.org/#1}{\path{#1}}}
\providecommand{\Pubmed}[1]{\href{pmid:#1}{\path{#1}}}
\providecommand{\bibinfo}[2]{#2}
\ifx\xfnm\relax \def\xfnm[#1]{\unskip,\space#1}\fi
%Type = Article
\bibitem[{Aldam et~al.(2017)Aldam, Weikamp, Spatschek, Brener and
  Bouchbinder}]{Aldam2017}
\bibinfo{author}{Aldam, M.}, \bibinfo{author}{Weikamp, M.},
  \bibinfo{author}{Spatschek, R.}, \bibinfo{author}{Brener, E.A.},
  \bibinfo{author}{Bouchbinder, E.}, \bibinfo{year}{2017}.
\newblock \bibinfo{title}{{Critical Nucleation Length for Accelerating
  Frictional Slip}}.
\newblock \bibinfo{journal}{Geophysical Research Letters} \bibinfo{volume}{44},
  \bibinfo{pages}{11,390--11,398}.
%\newblock \DOIprefix\doi{10.1002/2017GL074939}.
%Type = Article
\bibitem[{Ampuero and Rubin(2008)}]{Ampuero2008}
\bibinfo{author}{Ampuero, J.P.}, \bibinfo{author}{Rubin, A.M.}, \bibinfo{year}{2008}.
\newblock \bibinfo{title}{Earthquake nucleation on rate and state faults – {Aging} and slip laws}.
\newblock \bibinfo{journal}{Journal of Geophysical Research: Solid Earth} \bibinfo{volume}{113},
  \bibinfo{pages}{B01302}.
%\newblock \DOIprefix\doi{10.1029/2007JB005082}.
%Type = Article
\bibitem[{Bak et~al.(1988)Bak, Tang and Wiesenfeld}]{bak1988self}
\bibinfo{author}{Bak, P.}, \bibinfo{author}{Tang, C.},
  \bibinfo{author}{Wiesenfeld, K.}, \bibinfo{year}{1988}.
\newblock \bibinfo{title}{Self-organized criticality}.
\newblock \bibinfo{journal}{Physical Review A} \bibinfo{volume}{38},
  \bibinfo{pages}{364}.
%Type = Article
\bibitem[{Bar‐Sinai et~al.(2014)Bar‐Sinai, Spatschek, Brener and
  Bouchbinder}]{Barsinai2014}
\bibinfo{author}{Bar‐Sinai, Y.}, \bibinfo{author}{Spatschek, R.},
  \bibinfo{author}{Brener, E.A.}, \bibinfo{author}{Bouchbinder, E.},
  \bibinfo{year}{2014}.
\newblock \bibinfo{title}{On the velocity-strengthening behavior of dry
  friction}.
\newblock \bibinfo{journal}{Journal of Geophysical Research: Solid Earth}
  \bibinfo{volume}{119}, \bibinfo{pages}{1738--1748}.
%\newblock \DOIprefix\doi{10.1002/2013JB010586}.
%Type = Article
\bibitem[{Baumberger et~al.(1999)Baumberger, Berthoud and
  Caroli}]{Baumberger1999}
\bibinfo{author}{Baumberger, T.}, \bibinfo{author}{Berthoud, P.},
  \bibinfo{author}{Caroli, C.}, \bibinfo{year}{1999}.
\newblock \bibinfo{title}{Physical analysis of the state- and rate-dependent
  friction law. {II}. {Dynamic} friction}.
\newblock \bibinfo{journal}{Physical Review B} \bibinfo{volume}{60},
  \bibinfo{pages}{3928--3939}.
%\newblock \DOIprefix\doi{10.1103/PhysRevB.60.3928}.
%Type = Article
\bibitem[{Baumberger and Caroli(2006)}]{Baumberger2006}
\bibinfo{author}{Baumberger, T.}, \bibinfo{author}{Caroli, C.},
  \bibinfo{year}{2006}.
\newblock \bibinfo{title}{{Solid friction from stick–slip down to pinning and
  aging}}.
\newblock \bibinfo{journal}{Advances in Physics} \bibinfo{volume}{55},
  \bibinfo{pages}{279--348}.
%\newblock \DOIprefix\doi{10.1080/00018730600732186}.
%Type = Article
\bibitem[{Ben-David et~al.(2010)Ben-David, Rubinstein and
  Fineberg}]{Ben-david2010}
\bibinfo{author}{Ben-David, O.}, \bibinfo{author}{Rubinstein, S.M.},
  \bibinfo{author}{Fineberg, J.}, \bibinfo{year}{2010}.
\newblock \bibinfo{title}{Slip-stick and the evolution of frictional strength}.
\newblock \bibinfo{journal}{Nature} \bibinfo{volume}{463},
  \bibinfo{pages}{76--79}.
%\newblock \DOIprefix\doi{10.1038/nature08676}.
%Type = Article
\bibitem[{Ben-Zion(2001)}]{Ben-zion2001}
\bibinfo{author}{Ben-Zion, Y.}, \bibinfo{year}{2001}.
\newblock \bibinfo{title}{Dynamic ruptures in recent models of earthquake
  faults}.
\newblock \bibinfo{journal}{Journal of the Mechanics and Physics of Solids}
  \bibinfo{volume}{49}, \bibinfo{pages}{2209--2244}.
%\newblock \DOIprefix\doi{10.1016/S0022-5096(01)00036-9}.
%Type = Article
\bibitem[{Ben-Zion and Rice(1995)}]{Ben-zion1995}
\bibinfo{author}{Ben-Zion, Y.}, \bibinfo{author}{Rice, J.R.},
  \bibinfo{year}{1995}.
\newblock \bibinfo{title}{Slip patterns and earthquake populations along
  different classes of faults in elastic solids}.
\newblock \bibinfo{journal}{Journal of Geophysical Research: Solid Earth}
  \bibinfo{volume}{100}, \bibinfo{pages}{12959--12983}.
%\newblock \DOIprefix\doi{10.1029/94JB03037}.
%Type = Article
\bibitem[{Ben-Zion and Rice(1997)}]{ben1997dynamic}
\bibinfo{author}{Ben-Zion, Y.}, \bibinfo{author}{Rice, J.R.},
  \bibinfo{year}{1997}.
\newblock \bibinfo{title}{Dynamic simulations of slip on a smooth fault in an
  elastic solid}.
\newblock \bibinfo{journal}{Journal of Geophysical Research: Solid Earth}
  \bibinfo{volume}{102}, \bibinfo{pages}{17771--17784}.
%Type = Article
\bibitem[{Ben‐Zion and Rice(1993)}]{Ben-zion1993}
\bibinfo{author}{Ben‐Zion, Y.}, \bibinfo{author}{Rice, J.R.},
  \bibinfo{year}{1993}.
\newblock \bibinfo{title}{Earthquake failure sequences along a cellular fault
  zone in a three-dimensional elastic solid containing asperity and nonasperity
  regions}.
\newblock \bibinfo{journal}{Journal of Geophysical Research: Solid Earth}
  \bibinfo{volume}{98}, \bibinfo{pages}{14109--14131}.
%\newblock \DOIprefix\doi{10.1029/93JB01096}.
%Type = Article
\bibitem[{Bhattacharya and Rubin(2014)}]{Bhattacharya2014}
\bibinfo{author}{Bhattacharya, P.}, \bibinfo{author}{Rubin, A.M.},
  \bibinfo{year}{2014}.
\newblock \bibinfo{title}{{Frictional response to velocity steps and 1-D fault
  nucleation under a state evolution law with stressing-rate dependence}}.
\newblock \bibinfo{journal}{Journal of Geophysical Research: Solid Earth}
  \bibinfo{volume}{119}, \bibinfo{pages}{2272--2304}.
%\newblock \DOIprefix\doi{10.1002/2013JB010671}.
%Type = Book
\bibitem[{Bowden and Tabor(2001)}]{Bowden2001}
\bibinfo{author}{Bowden, F.P.}, \bibinfo{author}{Tabor, D.},
  \bibinfo{year}{2001}.
\newblock \bibinfo{title}{The {Friction} and {Lubrication} of {Solids}}.
\newblock \bibinfo{publisher}{Clarendon Press}.
%Type = Book
\bibitem[{Broberg(1999)}]{Broberg1999}
\bibinfo{author}{Broberg, K.B.}, \bibinfo{year}{1999}.
\newblock \bibinfo{title}{{Cracks and fracture}}.
\newblock \bibinfo{publisher}{Elsevier}.
%Type = Article
\bibitem[{Carlson and Langer(1989a)}]{Carlson1989b}
\bibinfo{author}{Carlson, J.M.}, \bibinfo{author}{Langer, J.S.},
  \bibinfo{year}{1989}a.
\newblock \bibinfo{title}{Mechanical model of an earthquake fault}.
\newblock \bibinfo{journal}{Physical Review A} \bibinfo{volume}{40},
  \bibinfo{pages}{6470--6484}.
%\newblock \DOIprefix\doi{10.1103/PhysRevA.40.6470}.
%Type = Article
\bibitem[{Carlson and Langer(1989b)}]{Carlson1989a}
\bibinfo{author}{Carlson, J.M.}, \bibinfo{author}{Langer, J.S.},
  \bibinfo{year}{1989}b.
\newblock \bibinfo{title}{Properties of earthquakes generated by fault
  dynamics}.
\newblock \bibinfo{journal}{Physical Review Letters} \bibinfo{volume}{62},
  \bibinfo{pages}{2632--2635}.
%\newblock \DOIprefix\doi{10.1103/PhysRevLett.62.2632}.
%Type = Article
\bibitem[{Carlson et~al.(1994)Carlson, Langer and Shaw}]{Carlson1994}
\bibinfo{author}{Carlson, J.M.}, \bibinfo{author}{Langer, J.S.},
  \bibinfo{author}{Shaw, B.E.}, \bibinfo{year}{1994}.
\newblock \bibinfo{title}{Dynamics of earthquake faults}.
\newblock \bibinfo{journal}{Reviews of Modern Physics} \bibinfo{volume}{66},
  \bibinfo{pages}{657--670}.
%\newblock \DOIprefix\doi{10.1103/RevModPhys.66.657}.
%Type = Article
\bibitem[{Carlson et~al.(1991)Carlson, Langer, Shaw and Tang}]{Carlson1991}
\bibinfo{author}{Carlson, J.M.}, \bibinfo{author}{Langer, J.S.},
  \bibinfo{author}{Shaw, B.E.}, \bibinfo{author}{Tang, C.},
  \bibinfo{year}{1991}.
\newblock \bibinfo{title}{Intrinsic properties of a {Burridge}-{Knopoff} model
  of an earthquake fault}.
\newblock \bibinfo{journal}{Physical Review A} \bibinfo{volume}{44},
  \bibinfo{pages}{884--897}.
%\newblock \DOIprefix\doi{10.1103/PhysRevA.44.884}.
%Type = Article
\bibitem[{Cattania(2019)}]{cattania2019complex}
\bibinfo{author}{Cattania, C.}, \bibinfo{year}{2019}.
\newblock \bibinfo{title}{Complex earthquake sequences on simple faults}.
\newblock \bibinfo{journal}{Geophysical Research Letters} \bibinfo{volume}{46},
  \bibinfo{pages}{10384--10393}.
%Type = Article
\bibitem[{Davidsen and Goltz(2004)}]{Davidsen2004}
\bibinfo{author}{Davidsen, J.}, \bibinfo{author}{Goltz, C.},
  \bibinfo{year}{2004}.
\newblock \bibinfo{title}{Are seismic waiting time distributions universal?}
\newblock \bibinfo{journal}{Geophysical Research Letters} \bibinfo{volume}{31},
  \bibinfo{pages}{L21612}.
%\newblock \DOIprefix\doi{10.1029/2004GL020892}.
%Type = Article
\bibitem[{Dieterich(1978)}]{Dieterich1978}
\bibinfo{author}{Dieterich, J.H.}, \bibinfo{year}{1978}.
\newblock \bibinfo{title}{Time-dependent friction and the mechanics of
  stick-slip}.
\newblock \bibinfo{journal}{Pure and Applied Geophysics} \bibinfo{volume}{116},
  \bibinfo{pages}{790--806}.
%\newblock \DOIprefix\doi{10.1007/BF00876539}.
%Type = Article
\bibitem[{Dieterich(1979)}]{Dieterich1979}
\bibinfo{author}{Dieterich, J.H.}, \bibinfo{year}{1979}.
\newblock \bibinfo{title}{Modeling of rock friction: 1. {Experimental} results
  and constitutive equations}.
\newblock \bibinfo{journal}{Journal of Geophysical Research: Solid Earth}
  \bibinfo{volume}{84}, \bibinfo{pages}{2161--2168}.
%\newblock \DOIprefix\doi{10.1029/JB084iB05p02161}.
%Type = Incollection
\bibitem[{Dieterich(1986)}]{Dieterich1986}
\bibinfo{author}{Dieterich, J.H.}, \bibinfo{year}{1986}.
\newblock \bibinfo{title}{A {Model} for the {Nucleation} of {Earthquake}
  {Slip}}, in: \bibinfo{booktitle}{Earthquake {Source} {Mechanics}}.
  \bibinfo{publisher}{American Geophysical Union (AGU)}, pp.
  \bibinfo{pages}{37--47}.
%\newblock \DOIprefix\doi{10.1029/GM037p0037}.
%Type = Article
\bibitem[{Dieterich(2007)}]{Dieterich2007}
\bibinfo{author}{Dieterich, J.H.}, \bibinfo{year}{2007}.
\newblock \bibinfo{title}{{Applications of rate-and state-dependent friction to
  models of fault slip and earthquake occurrence}}.
\newblock \bibinfo{journal}{Treatise Geophys.} \bibinfo{volume}{4},
  \bibinfo{pages}{107--129}.
%\newblock \DOIprefix\doi{10.1073/pnas.93.9.3787}.
%Type = Article
\bibitem[{Dieterich and Kilgore(1994)}]{Dieterich1994}
\bibinfo{author}{Dieterich, J.H.}, \bibinfo{author}{Kilgore, B.D.},
  \bibinfo{year}{1994}.
\newblock \bibinfo{title}{Direct observation of frictional contacts: {New}
  insights for state-dependent properties}.
\newblock \bibinfo{journal}{Pure and Applied Geophysics} \bibinfo{volume}{143},
  \bibinfo{pages}{283--302}.
%\newblock \DOIprefix\doi{10.1007/BF00874332}.
%Type = Article
\bibitem[{Fisher(1998)}]{fisher1998collective}
\bibinfo{author}{Fisher, D.S.}, \bibinfo{year}{1998}.
\newblock \bibinfo{title}{Collective transport in random media: from
  superconductors to earthquakes}.
\newblock \bibinfo{journal}{Physics Reports} \bibinfo{volume}{301},
  \bibinfo{pages}{113--150}.
%Type = Book
\bibitem[{Freund(1998)}]{freund1998dynamic}
\bibinfo{author}{Freund, L.B.}, \bibinfo{year}{1998}.
\newblock \bibinfo{title}{Dynamic fracture mechanics}.
\newblock \bibinfo{publisher}{Cambridge university press}.
%Type = Book
\bibitem[{Frisch(1995)}]{frisch1995turbulence}
\bibinfo{author}{Frisch, U.}, \bibinfo{year}{1995}.
\newblock \bibinfo{title}{Turbulence: the legacy of AN Kolmogorov}.
\newblock \bibinfo{publisher}{Cambridge university press}.
%Type = Article
\bibitem[{Gutenberg and Richter(1944)}]{Gutenberg1944}
\bibinfo{author}{Gutenberg, R.}, \bibinfo{author}{Richter, C.},
  \bibinfo{year}{1944}.
\newblock \bibinfo{title}{Frequency of earthquakes in {California}}.
\newblock \bibinfo{journal}{Bulletin of the Seismological Society of America}
  \bibinfo{volume}{34}, \bibinfo{pages}{185--188}.
%Type = Article
\bibitem[{Hansen et~al.(1991)Hansen, Hinrichsen and Roux}]{hansen1991roughness}
\bibinfo{author}{Hansen, A.}, \bibinfo{author}{Hinrichsen, E.L.},
  \bibinfo{author}{Roux, S.}, \bibinfo{year}{1991}.
\newblock \bibinfo{title}{Roughness of crack interfaces}.
\newblock \bibinfo{journal}{Physical Review Letters} \bibinfo{volume}{66},
  \bibinfo{pages}{2476}.
%Type = Article
\bibitem[{Horowitz and Ruina(1989)}]{horowitz1989slip}
\bibinfo{author}{Horowitz, F.G.}, \bibinfo{author}{Ruina, A.},
  \bibinfo{year}{1989}.
\newblock \bibinfo{title}{Slip patterns in a spatially homogeneous fault
  model}.
\newblock \bibinfo{journal}{Journal of Geophysical Research: Solid Earth}
  \bibinfo{volume}{94}, \bibinfo{pages}{10279--10298}.
%Type = Article
\bibitem[{Johnson(1992)}]{johnson1992influence}
\bibinfo{author}{Johnson, E.}, \bibinfo{year}{1992}.
\newblock \bibinfo{title}{The influence of the lithospheric thickness on
  bilateral slip}.
\newblock \bibinfo{journal}{Geophysical Journal International}
  \bibinfo{volume}{108}, \bibinfo{pages}{151--160}.
%Type = Article
\bibitem[{Kanamori and Anderson(1975)}]{Kanamori1975}
\bibinfo{author}{Kanamori, H.}, \bibinfo{author}{Anderson, D.L.},
  \bibinfo{year}{1975}.
\newblock \bibinfo{title}{Theoretical basis of some empirical relations in
  seismology}.
\newblock \bibinfo{journal}{Bulletin of the Seismological Society of America}
  \bibinfo{volume}{65}, \bibinfo{pages}{1073--1095}.
%\newblock \DOIprefix\doi{10.1785/BSSA0650051073}.
%Type = Article
\bibitem[{Kawamura et~al.(2012)Kawamura, Hatano, Kato, Biswas and
  Chakrabarti}]{Kawamura2012}
\bibinfo{author}{Kawamura, H.}, \bibinfo{author}{Hatano, T.},
  \bibinfo{author}{Kato, N.}, \bibinfo{author}{Biswas, S.},
  \bibinfo{author}{Chakrabarti, B.K.}, \bibinfo{year}{2012}.
\newblock \bibinfo{title}{Statistical physics of fracture, friction, and
  earthquakes}.
\newblock \bibinfo{journal}{Reviews of Modern Physics} \bibinfo{volume}{84},
  \bibinfo{pages}{839--884}.
%\newblock \DOIprefix\doi{10.1103/RevModPhys.84.839}.
%Type = Article
\bibitem[{Kilgore et~al.(1993)Kilgore, Blanpied and Dieterich}]{Kilgore1993}
\bibinfo{author}{Kilgore, B.D.}, \bibinfo{author}{Blanpied, M.L.},
  \bibinfo{author}{Dieterich, J.H.}, \bibinfo{year}{1993}.
\newblock \bibinfo{title}{Velocity dependent friction of granite over a wide
  range of conditions}.
\newblock \bibinfo{journal}{Geophysical Research Letters} \bibinfo{volume}{20},
  \bibinfo{pages}{903--906}.
%\newblock \DOIprefix\doi{10.1029/93GL00368}.
%Type = Article
\bibitem[{Langer et~al.(1996)Langer, Carlson, Myers and Shaw}]{Langer1996}
\bibinfo{author}{Langer, J.S.}, \bibinfo{author}{Carlson, J.M.},
  \bibinfo{author}{Myers, C.R.}, \bibinfo{author}{Shaw, B.E.},
  \bibinfo{year}{1996}.
\newblock \bibinfo{title}{Slip complexity in dynamic models of earthquake
  faults.}
\newblock \bibinfo{journal}{Proceedings of the National Academy of Sciences}
  \bibinfo{volume}{93}, \bibinfo{pages}{3825--3829}.
%\newblock \DOIprefix\doi{10.1073/pnas.93.9.3825}.
%Type = Article
\bibitem[{Lapusta and Liu(2009)}]{Lapusta2009_3D}
\bibinfo{author}{Lapusta, N.}, \bibinfo{author}{Liu, Y.}, \bibinfo{year}{2009}.
\newblock \bibinfo{title}{Three-dimensional boundary integral modeling of
  spontaneous earthquake sequences and aseismic slip}.
\newblock \bibinfo{journal}{Journal of Geophysical Research: Solid Earth}
  \bibinfo{volume}{114}, \bibinfo{pages}{B09303}.
%\newblock \DOIprefix\doi{10.1029/2008JB005934}.
%Type = Phdthesis
\bibitem[{Lebihain(2019)}]{lebihain2019}
\bibinfo{author}{Lebihain, M.}, \bibinfo{year}{2019}.
\newblock \bibinfo{title}{Large-scale crack propagation in heterogeneous
  materials: An insight into the homogenization of brittle fracture
  properties}.
\newblock Ph.D. thesis. Sorbonne Universit{\'e}.
\newblock \URLprefix \url{https://www.theses.fr/2019SORUS522}.
%Type = Article
\bibitem[{Lehner et~al.(1981)Lehner, Li and Rice}]{lehner1981stress}
\bibinfo{author}{Lehner, F.K.}, \bibinfo{author}{Li, V.C.},
  \bibinfo{author}{Rice, J.}, \bibinfo{year}{1981}.
\newblock \bibinfo{title}{Stress diffusion along rupturing plate boundaries}.
\newblock \bibinfo{journal}{Journal of Geophysical Research: Solid Earth}
  \bibinfo{volume}{86}, \bibinfo{pages}{6155--6169}.
%Type = Article
\bibitem[{Marone(1998)}]{Marone1998a}
\bibinfo{author}{Marone, C.}, \bibinfo{year}{1998}.
\newblock \bibinfo{title}{{Laboratoty-derived friction laws and their
  application to seismic faulting}}.
\newblock \bibinfo{journal}{Annual Review of Earth and Planetary Sciences}
  \bibinfo{volume}{26}, \bibinfo{pages}{643--696}.
%\newblock \DOIprefix\doi{10.1146/annurev.earth.26.1.643}.
%Type = Article
\bibitem[{Myers et~al.(1996)Myers, Shaw and Langer}]{Myers1996}
\bibinfo{author}{Myers, C.R.}, \bibinfo{author}{Shaw, B.E.},
  \bibinfo{author}{Langer, J.S.}, \bibinfo{year}{1996}.
\newblock \bibinfo{title}{Slip {Complexity} in a {Crustal}-{Plane} {Model} of
  an {Earthquake} {Fault}}.
\newblock \bibinfo{journal}{Physical Review Letters} \bibinfo{volume}{77},
  \bibinfo{pages}{972--975}.
%\newblock \DOIprefix\doi{10.1103/PhysRevLett.77.972}.
%Type = Article
\bibitem[{Nagata et~al.(2012)Nagata, Nakatani and Yoshida}]{Nagata2012}
\bibinfo{author}{Nagata, K.}, \bibinfo{author}{Nakatani, M.},
  \bibinfo{author}{Yoshida, S.}, \bibinfo{year}{2012}.
\newblock \bibinfo{title}{{A revised rate- and state-dependent friction law
  obtained by constraining constitutive and evolution laws separately with
  laboratory data}}.
\newblock \bibinfo{journal}{Journal of Geophysical Research: Solid Earth}
  \bibinfo{volume}{117}, \bibinfo{pages}{B02314}.
%\newblock \DOIprefix\doi{10.1029/2011JB008818}.
%Type = Article
\bibitem[{Nakatani(2001)}]{Nakatani2001}
\bibinfo{author}{Nakatani, M.}, \bibinfo{year}{2001}.
\newblock \bibinfo{title}{{Conceptual and physical clarification of rate and
  state friction: Frictional sliding as a thermally activated rheology}}.
\newblock \bibinfo{journal}{Journal of Geophysical Research: Solid Earth}
  \bibinfo{volume}{106}, \bibinfo{pages}{13347--13380}.
%\newblock \DOIprefix\doi{10.1029/2000JB900453}.
%Type = Article
\bibitem[{Nattermann et~al.(1992)Nattermann, Stepanow, Tang and
  Leschhorn}]{nattermann1992dynamics}
\bibinfo{author}{Nattermann, T.}, \bibinfo{author}{Stepanow, S.},
  \bibinfo{author}{Tang, L.H.}, \bibinfo{author}{Leschhorn, H.},
  \bibinfo{year}{1992}.
\newblock \bibinfo{title}{Dynamics of interface depinning in a disordered
  medium}.
\newblock \bibinfo{journal}{Journal de Physique II} \bibinfo{volume}{2},
  \bibinfo{pages}{1483--1488}.
%Type = Article
\bibitem[{Nielsen et~al.(2000)Nielsen, Carlson and Olsen}]{Nielsen2000}
\bibinfo{author}{Nielsen, S.B.}, \bibinfo{author}{Carlson, J.M.},
  \bibinfo{author}{Olsen, K.B.}, \bibinfo{year}{2000}.
\newblock \bibinfo{title}{Influence of friction and fault geometry on
  earthquake rupture}.
\newblock \bibinfo{journal}{Journal of Geophysical Research: Solid Earth}
  \bibinfo{volume}{105}, \bibinfo{pages}{6069--6088}.
%\newblock \DOIprefix\doi{https://doi.org/10.1029/1999JB900350}.
%Type = Article
\bibitem[{Omori(1894)}]{Omori1894}
\bibinfo{author}{Omori, F.}, \bibinfo{year}{1894}.
\newblock \bibinfo{title}{On the aftershocks of earthquakes}.
\newblock \bibinfo{journal}{Journal of the College of Science}
  \bibinfo{volume}{7}, \bibinfo{pages}{111--120}.
%Type = Article
\bibitem[{Reches and Lockner(2010)}]{Reches2010}
\bibinfo{author}{Reches, Z.}, \bibinfo{author}{Lockner, D.A.},
  \bibinfo{year}{2010}.
\newblock \bibinfo{title}{Fault weakening and earthquake instability by powder
  lubrication}.
\newblock \bibinfo{journal}{Nature} \bibinfo{volume}{467},
  \bibinfo{pages}{452--455}.
%\newblock \DOIprefix\doi{10.1038/nature09348}.
%Type = Article
\bibitem[{Rezakhani et~al.(2020)Rezakhani, Barras, Brun and
  Molinari}]{Rezakhani2020}
\bibinfo{author}{Rezakhani, R.}, \bibinfo{author}{Barras, F.},
  \bibinfo{author}{Brun, M.}, \bibinfo{author}{Molinari, J.F.},
  \bibinfo{year}{2020}.
\newblock \bibinfo{title}{Finite element modeling of dynamic frictional rupture
  with rate and state friction}.
\newblock \bibinfo{journal}{Journal of the Mechanics and Physics of Solids}
  \bibinfo{volume}{141}, \bibinfo{pages}{103967}.
%\newblock \DOIprefix\doi{10.1016/j.jmps.2020.103967}.
%Type = Article
\bibitem[{Rice(1993)}]{Rice1993}
\bibinfo{author}{Rice, J.R.}, \bibinfo{year}{1993}.
\newblock \bibinfo{title}{Spatio-temporal complexity of slip on a fault}.
\newblock \bibinfo{journal}{Journal of Geophysical Research: Solid Earth}
  \bibinfo{volume}{98}, \bibinfo{pages}{9885--9907}.
%\newblock \DOIprefix\doi{10.1029/93JB00191}.
%Type = Article
\bibitem[{Rice(North Holland Publishing Co., 1980)}]{rice1980mechanics}
\bibinfo{author}{Rice, J.R.}, \bibinfo{year}{North Holland Publishing Co.,
  1980}.
\newblock \bibinfo{title}{The mechanics of earthquake rupture}.
\newblock \bibinfo{journal}{In Physics of the Earth's Interior, Eds.
  A.M.~Dziewonski and E.~Boschi} .
%Type = Article
\bibitem[{Rice and Ben-Zion(1996)}]{Rice1996}
\bibinfo{author}{Rice, J.R.}, \bibinfo{author}{Ben-Zion, Y.},
  \bibinfo{year}{1996}.
\newblock \bibinfo{title}{Slip complexity in earthquake fault models.}
\newblock \bibinfo{journal}{Proceedings of the National Academy of Sciences}
  \bibinfo{volume}{93}, \bibinfo{pages}{3811--3818}.
%\newblock \DOIprefix\doi{10.1073/pnas.93.9.3811}.
%Type = Article
\bibitem[{Rice and Ruina(1983)}]{Rice1983}
\bibinfo{author}{Rice, J.R.}, \bibinfo{author}{Ruina, A.L.},
  \bibinfo{year}{1983}.
\newblock \bibinfo{title}{{Stability of Steady Frictional Slipping}}.
\newblock \bibinfo{journal}{Journal of Applied Mechanics} \bibinfo{volume}{50},
  \bibinfo{pages}{343--349}.
%\newblock \DOIprefix\doi{10.1115/1.3167042}.
%Type = Article
\bibitem[{Richart and Molinari(2015)}]{Richart2015}
\bibinfo{author}{Richart, N.}, \bibinfo{author}{Molinari, J.F.},
  \bibinfo{year}{2015}.
\newblock \bibinfo{title}{Implementation of a parallel finite-element library:
  {Test} case on a non-local continuum damage model}.
\newblock \bibinfo{journal}{Finite Elements in Analysis and Design}
  \bibinfo{volume}{100}, \bibinfo{pages}{41--46}.
%\newblock \DOIprefix\doi{10.1016/j.finel.2015.02.003}.
%Type = Article
\bibitem[{Roch et~al.(2022)Roch, Brener, Molinari and Bouchbinder}]{Roch2022}
\bibinfo{author}{Roch, T.}, \bibinfo{author}{Brener, E.A.},
  \bibinfo{author}{Molinari, J.F.}, \bibinfo{author}{Bouchbinder, E.},
  \bibinfo{year}{2022}.
\newblock \bibinfo{title}{Velocity-driven frictional sliding: {Coarsening} and
  steady-state pulses}.
\newblock \bibinfo{journal}{Journal of the Mechanics and Physics of Solids}
  \bibinfo{volume}{158}, \bibinfo{pages}{104607}.
%\newblock \DOIprefix\doi{10.1016/j.jmps.2021.104607}.
%Type = Article
\bibitem[{Rubino et~al.(2022)Rubino, Lapusta and Rosakis}]{Rubino2022}
\bibinfo{author}{Rubino, V.}, \bibinfo{author}{Lapusta, N.},
  \bibinfo{author}{Rosakis, A.J.}, \bibinfo{year}{2022}.
\newblock \bibinfo{title}{Intermittent lab earthquakes in dynamically weakening
  fault gouge}.
\newblock \bibinfo{journal}{Nature} \bibinfo{volume}{606},
  \bibinfo{pages}{922--929}.
%\newblock \DOIprefix\doi{10.1038/s41586-022-04749-3}.
%Type = Article
\bibitem[{Rubino et~al.(2017)Rubino, Rosakis and Lapusta}]{Rubino2017}
\bibinfo{author}{Rubino, V.}, \bibinfo{author}{Rosakis, A.J.},
  \bibinfo{author}{Lapusta, N.}, \bibinfo{year}{2017}.
\newblock \bibinfo{title}{Understanding dynamic friction through spontaneously
  evolving laboratory earthquakes}.
\newblock \bibinfo{journal}{Nature Communications} \bibinfo{volume}{8},
  \bibinfo{pages}{15991}.
%\newblock \DOIprefix\doi{10.1038/ncomms15991}.
%Type = Article
\bibitem[{Rubinstein et~al.(2004)Rubinstein, Cohen and
  Fineberg}]{Rubinstein2004}
\bibinfo{author}{Rubinstein, S.M.}, \bibinfo{author}{Cohen, G.},
  \bibinfo{author}{Fineberg, J.}, \bibinfo{year}{2004}.
\newblock \bibinfo{title}{Detachment fronts and the onset of dynamic friction}.
\newblock \bibinfo{journal}{Nature} \bibinfo{volume}{430},
  \bibinfo{pages}{1005--1009}.
%\newblock \DOIprefix\doi{10.1038/nature02830}.
%Type = Article
\bibitem[{Ruina(1983)}]{Ruina1983}
\bibinfo{author}{Ruina, A.L.}, \bibinfo{year}{1983}.
\newblock \bibinfo{title}{{Slip instability and state variable friction laws}}.
\newblock \bibinfo{journal}{Journal of Geophysal Research: Solid Earth}
  \bibinfo{volume}{88}, \bibinfo{pages}{10359--10370}.
%\newblock \DOIprefix\doi{10.1029/JB088iB12p10359}.
%Type = Article
\bibitem[{Rundle et~al.(2003)Rundle, Turcotte, Shcherbakov, Klein and
  Sammis}]{Rundle2003}
\bibinfo{author}{Rundle, J.B.}, \bibinfo{author}{Turcotte, D.L.},
  \bibinfo{author}{Shcherbakov, R.}, \bibinfo{author}{Klein, W.},
  \bibinfo{author}{Sammis, C.}, \bibinfo{year}{2003}.
\newblock \bibinfo{title}{Statistical physics approach to understanding the
  multiscale dynamics of earthquake fault systems}.
\newblock \bibinfo{journal}{Reviews of Geophysics} \bibinfo{volume}{41},
  \bibinfo{pages}{1019}.
%\newblock \DOIprefix\doi{10.1029/2003RG000135}.
%Type = Book
\bibitem[{Scholz(2019)}]{Scholz2019}
\bibinfo{author}{Scholz, C.H.}, \bibinfo{year}{2019}.
\newblock \bibinfo{title}{The {Mechanics} of {Earthquakes} and {Faulting}}.
\newblock \bibinfo{edition}{3} ed., \bibinfo{publisher}{Cambridge University
  Press}, \bibinfo{address}{Cambridge}.
%\newblock \DOIprefix\doi{10.1017/9781316681473}.
%Type = Article
\bibitem[{Sethna et~al.(2001)Sethna, Dahmen and Myers}]{sethna2001crackling}
\bibinfo{author}{Sethna, J.P.}, \bibinfo{author}{Dahmen, K.A.},
  \bibinfo{author}{Myers, C.R.}, \bibinfo{year}{2001}.
\newblock \bibinfo{title}{Crackling noise}.
\newblock \bibinfo{journal}{Nature} \bibinfo{volume}{410},
  \bibinfo{pages}{242--250}.
%Type = Article
\bibitem[{Shaw(1994)}]{Shaw1994}
\bibinfo{author}{Shaw, B.E.}, \bibinfo{year}{1994}.
\newblock \bibinfo{title}{Complexity in a spatially uniform continuum fault
  model}.
\newblock \bibinfo{journal}{Geophysical Research Letters} \bibinfo{volume}{21},
  \bibinfo{pages}{1983--1986}.
%\newblock \DOIprefix\doi{10.1029/94GL01685}.
%Type = Article
\bibitem[{Shaw(1997)}]{Shaw1997}
\bibinfo{author}{Shaw, B.E.}, \bibinfo{year}{1997}.
\newblock \bibinfo{title}{Model quakes in the two-dimensional wave equation}.
\newblock \bibinfo{journal}{Journal of Geophysical Research: Solid Earth}
  \bibinfo{volume}{102}, \bibinfo{pages}{27367--27377}.
%\newblock \DOIprefix\doi{10.1029/97JB02786}.
%Type = Article
\bibitem[{Shaw and Rice(2000)}]{shaw2000existence}
\bibinfo{author}{Shaw, B.E.}, \bibinfo{author}{Rice, J.R.},
  \bibinfo{year}{2000}.
\newblock \bibinfo{title}{Existence of continuum complexity in the
  elastodynamics of repeated fault ruptures}.
\newblock \bibinfo{journal}{Journal of Geophysical Research: Solid Earth}
  \bibinfo{volume}{105}, \bibinfo{pages}{23791--23810}.
%Type = Book
\bibitem[{Stauffer and Aharony(2018)}]{stauffer2018introduction}
\bibinfo{author}{Stauffer, D.}, \bibinfo{author}{Aharony, A.},
  \bibinfo{year}{2018}.
\newblock \bibinfo{title}{Introduction to percolation theory}.
\newblock \bibinfo{publisher}{CRC press}.
%Type = Article
\bibitem[{Tullis and Weeks(1986)}]{Tullis1986}
\bibinfo{author}{Tullis, T.E.}, \bibinfo{author}{Weeks, J.D.},
  \bibinfo{year}{1986}.
\newblock \bibinfo{title}{Constitutive behavior and stability of frictional
  sliding of granite}.
\newblock \bibinfo{journal}{Pure and Applied Geophysics} \bibinfo{volume}{124},
  \bibinfo{pages}{383--414}.
%\newblock \DOIprefix\doi{10.1007/BF00877209}.
%Type = Article
\bibitem[{Utsu(1999)}]{Utsu1999}
\bibinfo{author}{Utsu, T.}, \bibinfo{year}{1999}.
\newblock \bibinfo{title}{Representation and {Analysis} of the {Earthquake}
  {Size} {Distribution}: {A} {Historical} {Review} and {Some} {New}
  {Approaches}}.
\newblock \bibinfo{journal}{Pure and Applied Geophysics} \bibinfo{volume}{155},
  \bibinfo{pages}{509--535}.
%\newblock \DOIprefix\doi{10.1007/s000240050276}.
%Type = Article
\bibitem[{Utsu et~al.(1995)Utsu, Ogata, S and {Matsu'ura}}]{Utsu1995}
\bibinfo{author}{Utsu, T.}, \bibinfo{author}{Ogata, Y.}, \bibinfo{author}{S,
  R.}, \bibinfo{author}{{Matsu'ura}}, \bibinfo{year}{1995}.
\newblock \bibinfo{title}{The {Centenary} of the {Omori} {Formula} for a
  {Decay} {Law} of {Aftershock} {Activity}}.
\newblock \bibinfo{journal}{Journal of Physics of the Earth}
  \bibinfo{volume}{43}, \bibinfo{pages}{1--33}.
%\newblock \DOIprefix\doi{10.4294/jpe1952.43.1}.
%Type = Article
\bibitem[{Wells and Coppersmith(1994)}]{Wells1994}
\bibinfo{author}{Wells, D.L.}, \bibinfo{author}{Coppersmith, K.J.},
  \bibinfo{year}{1994}.
\newblock \bibinfo{title}{New empirical relationships among magnitude, rupture
  length, rupture width, rupture area, and surface displacement}.
\newblock \bibinfo{journal}{Bulletin of the Seismological Society of America}
  \bibinfo{volume}{84}, \bibinfo{pages}{974--1002}.
%\newblock \DOIprefix\doi{10.1785/BSSA0840040974}.
%Type = Article
\bibitem[{Weng and Ampuero(2019)}]{Weng2019}
\bibinfo{author}{Weng, H.}, \bibinfo{author}{Ampuero, J.P.},
  \bibinfo{year}{2019}.
\newblock \bibinfo{title}{The {Dynamics} of {Elongated} {Earthquake}
  {Ruptures}}.
\newblock \bibinfo{journal}{Journal of Geophysical Research: Solid Earth}
  \bibinfo{volume}{124}, \bibinfo{pages}{8584--8610}.
%\newblock \DOIprefix\doi{10.1029/2019JB017684}.
%Type = Article
\bibitem[{Weng and Yang(2017)}]{Weng2017}
\bibinfo{author}{Weng, H.}, \bibinfo{author}{Yang, H.}, \bibinfo{year}{2017}.
\newblock \bibinfo{title}{Seismogenic width controls aspect ratios of
  earthquake ruptures}.
\newblock \bibinfo{journal}{Geophysical Research Letters} \bibinfo{volume}{44},
  \bibinfo{pages}{2725--2732}.
%\newblock \DOIprefix\doi{10.1002/2016GL072168}.
%Type = Article
\bibitem[{Zaiser(2006)}]{zaiser2006scale}
\bibinfo{author}{Zaiser, M.}, \bibinfo{year}{2006}.
\newblock \bibinfo{title}{Scale invariance in plastic flow of crystalline
  solids}.
\newblock \bibinfo{journal}{Advances in Physics} \bibinfo{volume}{55},
  \bibinfo{pages}{185--245}.

\end{thebibliography}

\end{document}